%% file: main.tex
\documentclass[apj,tighten,iop]{emulateapj2}

\usepackage{float}
\usepackage{graphicx}

\graphicspath{{/Users/aberti/Desktop/Research/clustering/ms/figures/}}
\makeatletter
\def\input@path{{/Users/aberti/Desktop/Research/clustering/ms/tables/}}
\makeatother

\input{defs}

\begin{document}
\title{PRIMUS:~Clustering of Star-forming and Quiescent Central Galaxies at $0.2 < {\rm \MakeLowercase{\it z}} < 0.9$}
\shorttitle{sSFR Dependence of Central Galaxy Clustering}
\shortauthors{Berti et al.}

\author{Angela M. Berti\altaffilmark{1},
	Alison L. Coil\altaffilmark{1},
	Andrew P. Hearin\altaffilmark{2},
	John Moustakas\altaffilmark{3}
	}
	
\altaffiltext{1}{Center for Astrophysics and Space Sciences, Department of Physics, University of California, 9500 Gilman Dr., La Jolla, San Diego, CA 92093, USA}
\altaffiltext{2}{High-Energy Physics Division, Argonne National Laboratory, Argonne, IL 60439, USA}
\altaffiltext{3}{Department of Physics and Astronomy, Siena College, 515 Loudon Road, Loudonville, NY 12211, USA}


\begin{abstract}

Previous work has demonstrated that at a given stellar mass, quiescent galaxies are more strongly clustered than star-forming galaxies. The contribution to this signal from central, as opposed to satellite,  galaxies is not known, which has strong implications for galaxy evolution models.
To investigate the contribution from central galaxies, here we present measurements of the clustering of isolated primary (IP) galaxies, used as a proxy for central galaxies, at ${0.2<z<0.9}$ with data from the PRIMUS galaxy redshift survey.
Using a sample of spectroscopic redshifts for $\sim60,000$ galaxies with ${M_*\gtrsim10^9\msun}$ covering 5~deg$^2$ on the sky, we define IP galaxies using isolation cuts in spatial proximity and stellar mass  of nearby galaxies.
We find that at fixed stellar mass, quiescent IP galaxies are more strongly clustered than star-forming IP galaxies at $z\sim0.35$ $(10\sigma)$.
Using mock galaxy catalogs based on recent halo occupation models of \cite{behroozi_etal18} and designed to replicate the parameters of the PRIMUS survey dataset, we find that these clustering differences are due in part to quiescent  central galaxies being more strongly clustered than star-forming central galaxies.
This is consistent with either distinct stellar-to-halo mass relations for quiescent and star-forming central galaxies, and/or central galaxy assembly bias.
We additionally use mock catalogs to assess the dependence of both incompleteness and satellite galaxy contamination in the IP galaxy samples on redshift, galaxy type, and stellar mass, and demonstrate how isolation criteria yield biased subsamples of central galaxies via {\it environmental incompleteness}, or the preferential exclusion of central galaxies in overdense environments.
\end{abstract}

\input{intro}

\input{data}

\input{mocks}

\input{methods}

\input{results_clustering}

\input{results_comp_contam}

\input{conclusion}

\acknowledgements
This paper includes data gathered with the 6.5 m Magellan Telescopes located at Las Campanas Observatory, Chile.
We thank the support staff at LCO for their help during our observations, and we acknowledge the use of community access through NOAO observing time.
Funding for PRIMUS has been provided by NSF grants AST-0607701, 0908246, 0908442, 0908354, and NASA grant 08-ADP08-0019.
Special thanks to Peter Behroozi for help with generating mock catalogs with UniverseMachine, and to the PRIMUS survey team.
AMB and ALC acknowledge support from the Ingrid and Joseph W.~Hibben endowed chair at UC San Diego.

\bibliography{references}

\end{document}

%% file: defs.tex
\usepackage{natbib}
\usepackage[usenames]{color}
\definecolor{red}{rgb}{0.75,0,0}
\definecolor{green}{rgb}{0,0.5,0}
\definecolor{blue}{rgb}{0,0,0.75}
\usepackage[ colorlinks=true, 
             urlcolor=blue,
             filecolor=blue,
             linkcolor=red,
             citecolor=blue,
             pdftitle={pdf title},
             pdfauthor={pdf author},
             pagebackref, 
             bookmarks,
             bookmarksopen=true]{hyperref}

\usepackage{afterpage}
\usepackage{xspace}    
\usepackage{microtype} 
\usepackage{amsmath}   
\usepackage{amssymb}

\usepackage[normalem]{ulem}
\newcommand\redline{\bgroup\markoverwith{\textcolor{red}{\rule[0.5ex]{2pt}{0.4pt}}}\ULon}

\usepackage{multirow}
\usepackage{makecell} 
\usepackage{rotating} 

\newcommand{\degsq}{\ensuremath{\mathrm{deg}^2}}
\newcommand{\msun}{\ensuremath{{M}_{\odot}}}

\newcommand{\logm}{\ensuremath{\log({M}_*/{M}_{\odot})}}
\newcommand{\mstar}{\ensuremath{{M}_*}\xspace}
\newcommand{\mfrac}{\ensuremath{m_{\rm frac}}\xspace}
\newcommand{\mlim}{\ensuremath{{M}_{\rm lim}}\xspace}

\newcommand{\mIP}{\ensuremath{{M_{\text{IP}}}}\xspace}

\newcommand{\kms}{km~s$^{-1}$\xspace}
\newcommand{\sigmaz}{\ensuremath{\sigma_z}\xspace}

\newcommand{\xirppi}{\ensuremath{\xi(r_{\rm p},\,\pi)}\xspace}
\newcommand{\pimax}{\ensuremath{\pi_\textrm{max}}}

\newcommand{\iSEDfit}{\texttt{iSEDfit}\xspace}
\newcommand{\IDL}{\texttt{IDL}\xspace}

\newcommand{\maketmp}[1]{\expandafter\MakeUppercase\expandafter{#1}}

\newcommand{\sfrunit}{\ensuremath{\msun\,\textrm{yr}^{-1}}}

\renewcommand{\wp}{\ensuremath{w_{\rm p}}\xspace}
\newcommand{\wprp}{\ensuremath{w_{\rm p}(r_{\rm p})}\xspace}

\newcommand{\xir}{\ensuremath{\xi(r)}\xspace}
\newcommand{\rp}{\ensuremath{r_{\rm p}}\xspace}

\newcommand{\zmock}{\ensuremath{z_{\rm mock}}\xspace}

%% file: intro.tex
\section{Introduction}\label{sec:intro}

In the current $\Lambda$CDM paradigm, galaxies form at the centers of overdense regions where density fluctuations in the early universe have collapsed to form 
dark matter halos \citep{white_rees78}. The assembly of halos and their clustering properties can be modeled across cosmic time with cosmological simulations, while the clustering of galaxies is measured using large galaxy surveys. Theoretical models for how galaxies populate halos thus provide a bridge between observations of galaxy clustering and simulations of dark matters halos. 

Historically, the first statistical models for the galaxy--halo connection were predicated upon the assumption that present-day halo mass determines the galaxy content of a halo \citep[e.g.][]{peacock_smith00, seljak_etal00, berlind_weinberg02}. This assumption has proven to be remarkably powerful. Subsequent refinements of these models, such as the Halo Occupation Distribution (HOD), the Conditional Luminosity Function (CLF), and abundance matching have shown that a wide variety of large-scale structure measurements are consistent with the existence of a tight scaling relation between central galaxy stellar mass (or luminosity) and host halo mass \citep{tinker_etal05, zehavi_etal05, yang_etal03, coil_etal06, cacciato_etal13, kravtsov_etal04, conroy_etal06, wake_etal11, reddick_etal13, leauthaud_etal12}. Moreover, in the HOD and CLF, the total number of satellite galaxies brighter than some threshold scales simply as a power law with  halo mass.

In addition to the above trends based on stellar mass or luminosity, it has been known for many years that two-point clustering and weak lensing has additional, strong dependence upon broadband color 
\citep[e.g.,][and references therein]{coil_etal08, zehavi_etal11, mandelbaum_etal16}. 
To capture these trends, models of the HOD and CLF have been extended such that more massive dark matter halos host larger fractions of quenched (red) galaxy populations \citep[e.g.,][]{vdb03}. This basic modeling assumption has been strikingly successful at fitting the color-dependence of two-point clustering \citep{zehavi_etal11}, galaxy-galaxy lensing \citep{tinker_etal13, zu_mandelbaum16b}, and a wide variety of other measurements \citep[for a recent review see][]{wechsler_tinker18}.

While galaxy clustering dependencies historically have been demonstrated in broad bins of luminosity, stellar mass, or color, recent work has shown that the dependence of galaxy clustering on specific star formation rate (sSFR, or SFR per unit stellar mass) remains strong when measured at {\it fixed} stellar mass, both at $z\sim0.1$ \citep{watson_etal15} and at higher redshift with the PRIMUS survey \citep{coil_etal17}.
Additionally, weak gravitational lensing studies have found that at fixed stellar mass, red quiescent galaxies reside in more massive halos than blue star-forming galaxies \citep{velander_etal13, rodriguez-puebla_etal15, mandelbaum_etal16}.

As models of the relationship between galaxies and their halos continue to be refined, a natural question is whether, and to what extent, various galaxy properties depend on properties of halos besides mass. This question first arose after the discovery of so-called ``halo assembly bias": at fixed halo mass, the clustering of simulated halos shows strong dependence on halo formation time \citep{gao_etal05}, halo concentration, and other properties \citep{wechsler_etal06, dalal_etal08, villarreal17, mao_etal18, salcedo_etal18, johnson_etal18, mansfield_kravtsov19}. Thus if the true statistical connection between galaxies and halos has additional dependence on halo assembly, then the standard ``halo mass only" assumption of the HOD can lead to misinterpretation of galaxy clustering measurements, a phenomenon referred to as ``galaxy assembly bias" \citep{zentner_etal16, wechsler_tinker18}.

The relationship between galaxy clustering and sSFR was largely unexplored until recently \citep[e.g.][]{watson_etal15}.
\citet{coil_etal17} measured the dependence of galaxy clustering on sSFR at fixed stellar mass to $z\sim1.2$ with the PRIMUS and DEEP2 spectroscopic galaxy redshift surveys, and found that at a given stellar mass quiescent (low sSFR) galaxies are more strongly clustered than star-forming (high sSFR) galaxies.
They also showed that within each of the star-forming and quiescent galaxy populations, galaxies with lower sSFR are more strongly clustered, at a given stellar mass.  Their results comparing the clustering dependence on both sSFR and stellar mass imply that clustering depends as strongly on sSFR as it does on stellar mass.

It is not yet known whether the observed correlation between clustering strength and sSFR is primarily due to central or satellite galaxies, or a combination of both.
The strong clustering dependence with sSFR reported in \citet{coil_etal17} could, in principle, be entirely due to satellite galaxies: relative to central galaxies of the same stellar mass, satellites have a larger quiescent fraction \citep{wetzel_etal12} and reside in more massive halos \citep{watson_conroy13}. This degeneracy highlights the potential constraining power of robust and precise measurements of {\em central} galaxy clustering. Resolving the sSFR dependence of galaxy clustering on central versus satellite galaxies has strong implications for halo occupation models, as it could imply that the stellar-to-halo mass relation depends on galaxy sSFR.

Here, to investigate the extent to which the correlation between clustering strength and sSFR at a given stellar mass may exist for central galaxies, we measure the relative bias (i.e., the ratio of clustering amplitudes) on two-halo scales ($1<\rp<10$~Mpc$/h$) of quiescent and star-forming ``isolated primary" (IP) galaxies at ${0.2<z<0.9}$ with data from the PRIMUS galaxy redshift survey. An IP galaxy has no other galaxies above a given stellar mass threshold within a cylindrical volume specified by a projected radius and line-of-sight distance, and are commonly used as an observational proxy for central galaxies. We also measure the clustering and relative bias of star-forming (quiescent) IP galaxies above and below the star-forming (quiescent) sequence at ${0.2<z<0.7}$ to probe whether clustering amplitude depends on sSFR for central galaxies \emph{within} each sequence. Using PRIMUS-like mock catalogs based on UniverseMachine \citep{behroozi_etal18}, we carefully scrutinize potential biases that may be caused by (i) satellite galaxies contaminating our IP galaxy samples, and (ii) systematically incomplete sampling of the distribution of large-scale environments of true central galaxies, a phenomenon we dub {\em environmental incompleteness.}

The structure of this paper is as follows. In \S\ref{sec:data} we present details of the PRIMUS spectroscopic redshift survey data used here and summarize the properties of our PRIMUS IP galaxy samples.
\S\ref{sec:mocks} presents details of our PRIMUS-like mock galaxy catalogs and the mock galaxy samples.
The methods we use to measure clustering and relative galaxy bias are described in \S\ref{sec:methods}, and the results presented in \S\ref{sec:results_clustering}.
In \S\ref{sec:results_comp_contam} we quantify the completeness and contamination of our galaxy samples, and discuss the implications of these systematic sources of error on studies that utilize isolation criteria to select central galaxy samples.
We summarize our main results and discuess their implications in \S\ref{sec:summary}.
Throughout this paper we assume a standard $\Lambda$CDM cosmology with $\Omega_m=0.3$, $\Omega_{\Lambda}=0.7$, and $H_0=70$~km~s$^{-1}$~Mpc$^{-1}$.

%% file: data.tex
\section{Data and Galaxy Samples}\label{sec:data}

In this section we describe the PRIMUS redshift survey data and the mock galaxy catalogs we create to compare with the clustering results from PRIMUS. We describe how we identify star-forming and quiescent galaxies and how we define isolated primary galaxies in both the PRIMUS dataset and the mock catalogs.

\begin{figure*}[tb]
\centering
\includegraphics[width=0.8\linewidth]{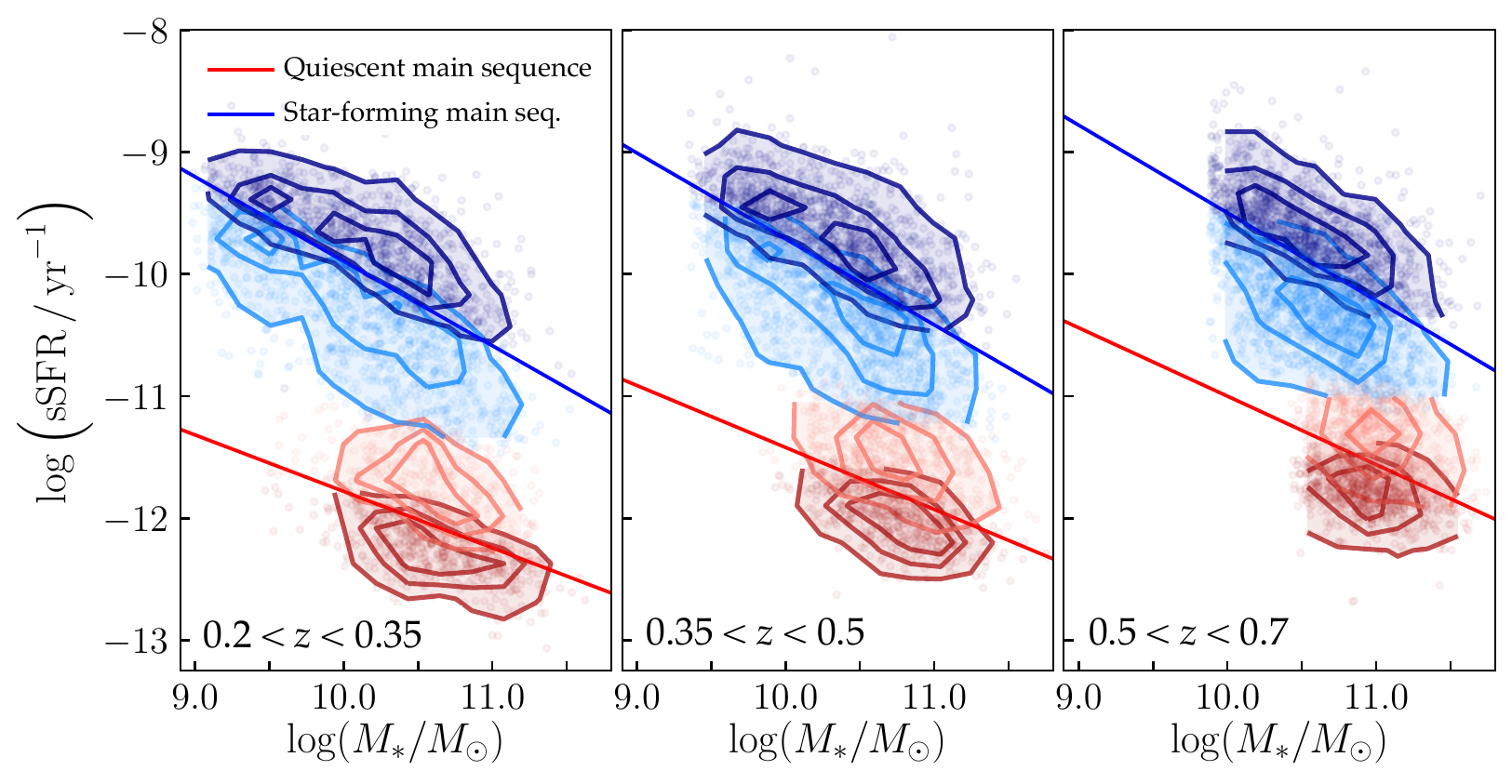}
\caption{
Stellar mass versus specific star formation rate (sSFR) for the four PRIMUS main sequence split IP galaxy samples (see \S\ref{sec:ms_split}) in three redshift bins:~${0.2 < z < 0.35}$ (left), ${0.35 < z < 0.5}$ (middle), and ${0.5 < z < 0.7}$ (right).
Galaxies are divided into star-forming (blue) and quiescent (red) populations via Equation~\ref{eq:sfr_cut}, a cut in the stellar mass-SFR plane that evolves with redshift and intersects the minimum of the bimodal galaxy distribution.
Solid blue and red lines in each panel show the cuts dividing IP galaxies into samples above (SF-above; dark blue) and below (SF-below; light blue) the star-forming main sequence (Eq.~\ref{eq:primus_ms_cut_blue}) and above (Q-above; light red) and below (Q-below; dark red) the quiescent main sequence (Eq.~\ref{eq:primus_ms_cut_red}) evaluated at the mean redshift of that panel.
}
\label{fig:primus_ms_cuts}
\end{figure*}

\begin{figure}[b]
\centering
\includegraphics[trim={0.18cm 0 0 0},clip,width=\linewidth]{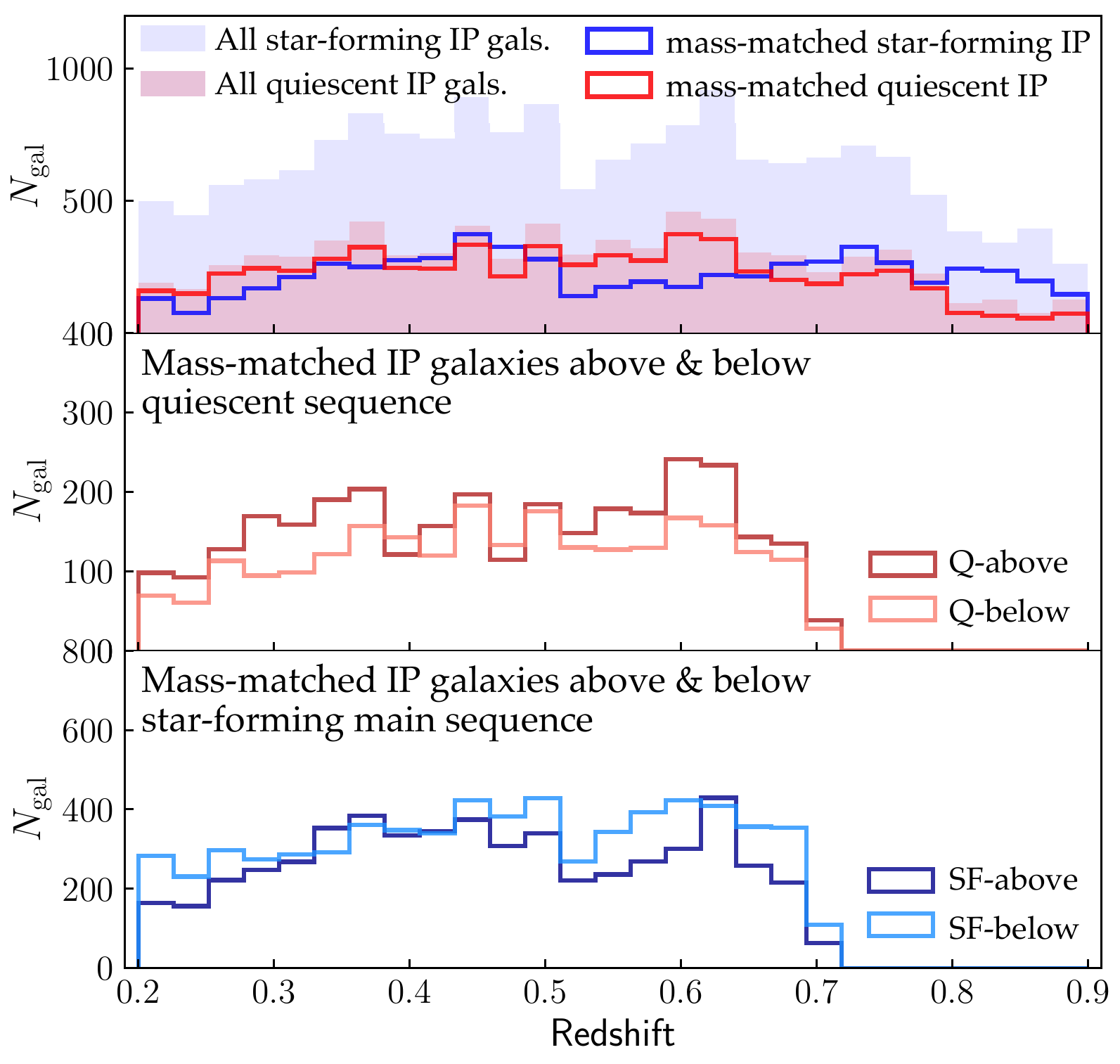}
\caption{
Top panel:~Filled histograms show the redshift distributions of {\em all} quiescent (red) and star-forming (blue) PRIMUS isolated primary (IP; see \S\ref{sec:ip_selection} for details) galaxies from $0.2<z<0.9$. Unfilled histograms show stellar mass and redshift-matched quiescent and star-forming IP galaxy samples, each of which is a subset of the corresponding filled histogram samples.
Middle panel:~Stellar mass-matched quiescent PRIMUS IP galaxies divided into samples above (light red; ``Q-above") and below (dark red; ``Q-below") the quiescent main sequence (see \S\ref{sec:ms_split} for details).
Bottom panel:~Stellar mass-matched star-forming PRIMUS IP galaxies divided into samples above (dark blue; ``SF-above") and below (light blue; ``SF-below") the star-forming main sequence.
}
\label{fig:primus_redshift_hist}
\end{figure}

\begin{figure}[b]
\centering
\includegraphics[width=\linewidth]{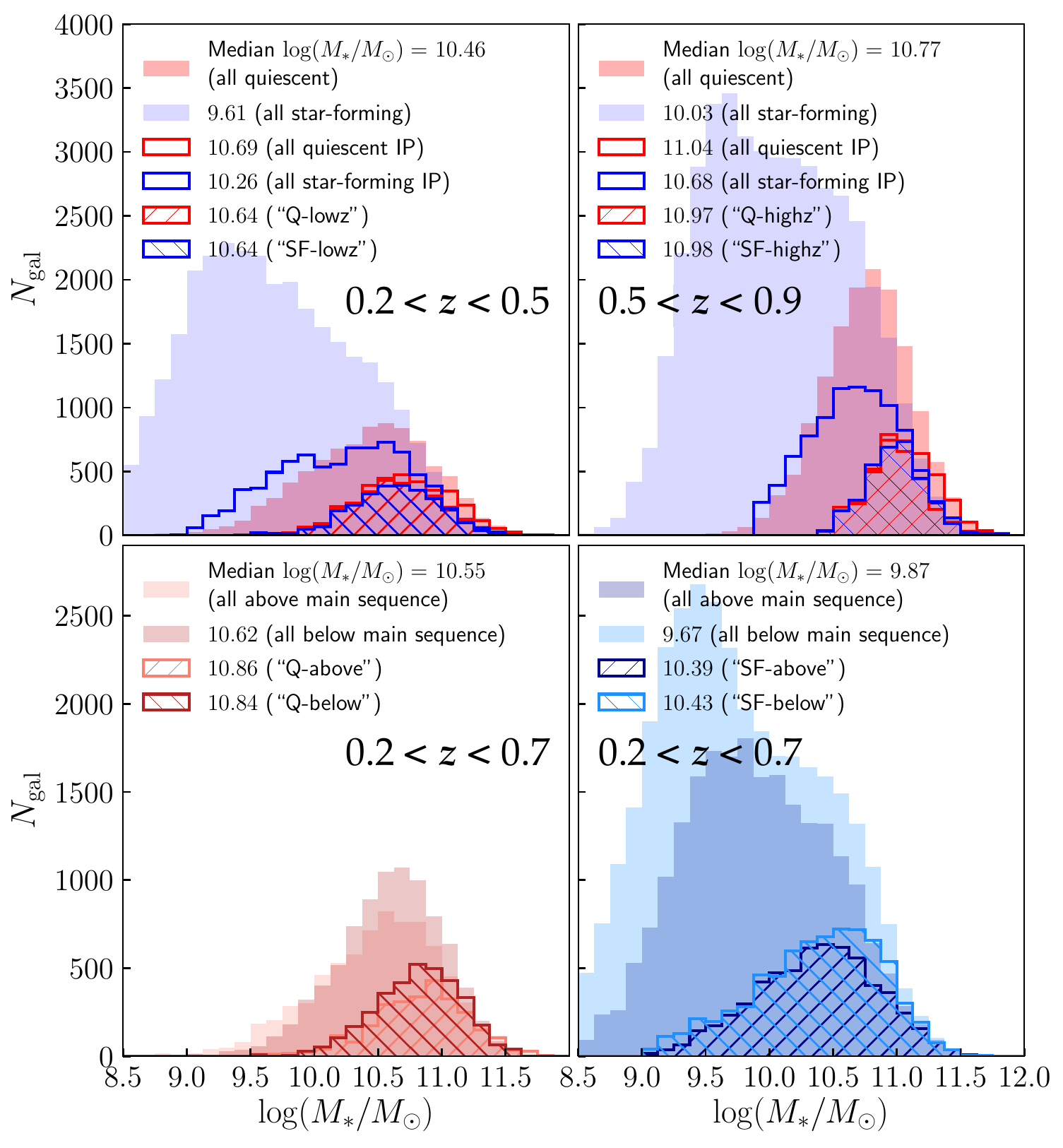}
\caption{
Top panels:~Stellar mass distributions of all star-forming and quiescent PRIMUS galaxies (filled blue and red histograms), all IP galaxies (unfilled blue and red histograms), and stellar mass-matched distributions of star-forming and quiescent IP galaxies (``star-forming/quiescent split" samples; hatched blue and red histograms) in low ($0.2<z<0.5$; left) and high ($0.5<z<0.9$; right) redshift bins.
Bottom panels:~Filled histograms show the stellar mass distributions of all PRIMUS galaxies above and below the quiescent (left) and star-forming (right) main sequences at $0.2<z<0.7$. Hatched histograms show the stellar mass distributions of corresponding PRIMUS IP galaxy samples (``main sequence split" samples).
Also shown is the median stellar mass of each galaxy sample.
}
\label{fig:primus_mass_hist}
\end{figure}

\subsection{The PRIMUS Redshift Survey}\label{sec:PRIMUS}
 
The PRIsm MUlti-Object Survey (PRIMUS) is the largest spectroscopic faint galaxy redshift survey completed to date.
The survey was conducted with the IMACS spectrograph \citep{bigelow_dressler03} on the Magellan I Baade 6.5-meter telescope at Las Campanas Observatory, using slitmasks and a low-dispersion prism.
The design allowed for $\sim2,000$ objects per slitmask to be observed simultaneously with a spectral resolution of ${\lambda/\Delta
\lambda \sim 40}$ in a $\sim0.2~\degsq$ field of view.
Objects were targeted to a maximum depth of ${i \ge 23}$, and typically two slitmasks were observed per pointing on the sky.  
PRIMUS obtained robust redshifts \citep[$Q\ge3$;~see][]{cool_etal13} for $\sim120,000$ objects at ${0<z<1.2}$ with a redshift precision of ${\sigmaz/(1 + z) \sim 0.005}$.

The total survey area of PRIMUS is $9.1~\degsq$ and encompasses seven distinct science fields.
Here we use the PRIMUS fields that have deep multi-wavelength ultraviolet (UV) imaging from the Galaxy Evolution Explorer 
\citep[GALEX;][]{martin_etal05}, 
mid-infrared imaging from the Spitzer Space Telescope \citep{werner_etal04} Infrared Array Camera \citep[IRAC;][]{fazio_etal04}, and optical and 
near-IR imaging 
from various ground-based surveys.
These include 
the Chandra Deep Field South-SWIRE field \citep[CDFS;][]{lonsdale_etal03},
the COSMOS field \citep{scoville_etal07},
the European Large Area ISO Survey-South 1 field \citep[ES1;][]{oliver_etal00},
and two spatially-adjacent subfields of the XMM-Large Scale Structure Survey field \citep[XMM-LSS;][]{pierre_etal04}.
The XMM subfields are the Subaru/XMM-Newton DEEP Survey field \citep[XMM-SXDS;][]{furusawa_etal08} and the Canada-France-Hawaii 
Telescope Legacy 
Survey (CFHTLS) field (XMM-CFHTLS).
The data used here cover a total of $\sim5.5~\degsq$ on the sky.

Full details of the survey design, targeting, and data summary can be found in \citet{coil_etal11}, while details of data reduction, redshift 
fitting, precision, 
and survey completeness are available in \citet{cool_etal13}.

\subsection{Full Galaxy Sample and Targeting Weights}\label{sec:targ_weight}

Objects in PRIMUS are classified as galaxies, stars, or broad-line AGN by fitting the low-resolution spectra and multi-wavelength photometry 
for each source with an empirical library of templates.
The best-fit template defines both the redshift and the type of the source.  
We exclude AGN from this study, where the optical light is dominated by the AGN and we therefore can not determine host properties.  We keep only those objects defined as galaxies with robust redshifts {($Q\ge 3$)} in the redshift range ${0.2<z<1.0}$.
We also only use galaxies with well-defined targeting weights (these are termed ``primary'' galaxies in \citet{coil_etal11}; we do not use that naming here, to avoid confusion with our isolated primary samples defined below in \S\ref{sec:gal_samples}).
These galaxies have a well-understood spatial and slitmask targeting selection function, defined by both a density-dependent weight and a magnitude-dependent sparse-sampling 
weight. In combination with a third, post-targeting weight that accounts for redshift incompleteness (see below) these weights allow a statistically complete galaxy sample to be recovered, which is suitable for analysis on two-point statistics, such as performed here.
PRIMUS targeting weights are described in detail in \citet{coil_etal11} and \citet{cool_etal13}.
Briefly, density-dependent weights account for sources that PRIMUS could not target in dense survey regions, as galaxies are sufficiently clustered in the plane of
the sky to the PRIMUS flux limit that even two slitmasks per pointing could not target every galaxy below the magnitude limit in each field (as spectra would overlap on the detector). 
Sparse-sampling weights are magnitude-dependent and ensure that the PRIMUS target catalog is not dominated by the faintest objects within the survey flux limit. Sparse-sampling weights were used to randomly select roughly a third 
of galaxies in the faintest 0.5~mag interval above the primary sample targeting limit.

\citet{skibba_etal14} measured galaxy clustering in PRIMUS and tested the recoverability of two-point statistics with the PRIMUS dataset using mock galaxy catalogs.
They applied the same process used to select the PRIMUS target sample and calculate  density-dependent and sparse-sampling weights to a mock catalog, and generated a weighted mock sample. 
\citet{skibba_etal14} then compared the correlation function of all galaxies in the mock catalog to that of the weighted mock sample and found no systematic difference between the two. Thus when PRIMUS targeting weights are applied to the survey data, two-point statistics can be accurately recovered.

A third, post-targeting weight \citep[described in detail in][]{cool_etal13} accounts for the fact that not all PRIMUS spectra yielded reliable {($Q \ge 3$)} redshifts.
As shown in \S7 of \citet{cool_etal13}, the PRIMUS redshift success rate is primarily a function of $i$-band magnitude and does \emph{not} depend strongly on galaxy color. 
Taken together, the three weights described above allow for the recovery of a statistically complete galaxy sample from the targeted sources with reliable redshifts.

\subsection{Stellar Mass and SFR Estimates}\label{sec:SFR}

Stellar masses and star formation rates (SFRs) for PRIMUS galaxies are obtained using SED fitting, a widely-adopted method for estimating the physical properties of galaxies.
A complete description of the \iSEDfit fitting process used here can be found in \citet{moustakas_etal13}, and we summarize the relevant points here.

\iSEDfit is a suite of routines written in the \IDL programming language that uses galaxy redshifts and photometry to compute the statistical likelihood of a large ensemble of model SEDs for each galaxy.
Model SEDs are generated using population synthesis models that assume a universal \citet{chabrier03} initial mass function (IMF) from $0.1-100~\msun$, and span a wide range of observed colors and physical properties (age, metallicity, star formation history, dust content, etc.).
\iSEDfit uses a Monte Carlo technique to randomly select values of model parameters from user-defined parameter distributions and compute a posterior probability distribution function (PDF).
PDFs of stellar mass and SFR are found by marginalizing over all other parameters, and the median value of the marginalized PDF is taken as the best estimate of the stellar mass or SFR of each galaxy.

\citet{berti_etal17} tested how the uncertainties on the stellar mass and SFR estimates described above affect the classification of galaxies as either star-forming or quiescent by randomly sampling individual stellar masses and SFRs for each galaxy in the full sample 100 times from normal distributions with widths equal to the stellar mass or SFR error for that galaxy, and found is an average change in the star-forming fraction of $<1\%$.

\subsection{Identifying Star-forming and Quiescent Galaxies}\label{sec:sfq}

To divide our sample into star-forming and quiescent galaxies we use a cut in the SFR\textendash stellar mass plane that evolves with redshift. This cut traces the minimum of the bimodal PRIMUS galaxy distribution seen in this plane, in six redshift bins  from ${z=0.2}$--1.0 \citep[see Figure 2 of][]{berti_etal17}, and is given by the following linear relation:
\begin{equation}
\log({\rm SFR}) = -1.29 + 0.65\log(\mstar - 10) + 1.33(z - 0.1)
\label{eq:sfr_cut}
\end{equation}

\noindent where SFR has units of \sfrunit and \mstar has units of \msun.
The slope of this line is defined by the slope of the star-forming main sequence \citep[e.g.,][]{noeske_etal07} as measured in the PRIMUS dataset using \iSEDfit SFR and stellar mass estimates. 
Each galaxy is classified as star-forming or quiescent based on whether it lies above or below the cut defined by Equation~\ref{eq:sfr_cut}, evaluated at the redshift of the galaxy.
Figure~\ref{fig:primus_ms_cuts} shows stellar mass versus specific SFR for PRIMUS galaxies in three redshift bins between ${z=0.2}$ and ${z=0.7}$. Blue and red contours represent star-forming and quiescent galaxies, respectively. The light and dark shades of red and blue are described in \S\ref{sec:ms_split} below.

\subsection{Stellar Mass Completeness Limits}\label{sec:smcl}

Because PRIMUS is a flux-limited survey targeted in the $i$ band, star-forming galaxies can be more easily detected at lower stellar mass than quiescent galaxies.
This results in a disproportionate number of star-forming galaxies at lower stellar masses.
To account for this we define a stellar mass limit above which at least 95\% of all galaxies can be detected, regardless of their SFR.
This stellar mass completeness limit is a function of redshift, galaxy type (star-forming or quiescent), and also varies slightly between fields (due to the different photometry used for targeting in each field).
Details of the calculation of PRIMUS mass completeness limits can be found in \citet{moustakas_etal13}.
When selecting isolated primary (IP) galaxies (see \S\ref{sec:ip_selection} below) we only consider a galaxy to be an IP candidate if its stellar mass is at least twice the mass completeness limit for that galaxy. This ensures we avoid misclassifying as isolated galaxies that \emph{may} have one or more sufficiently massive yet undetected neighboring galaxies such that the IP candidate should not be considered isolated.

\subsection{Isolated Primary Selection}\label{sec:ip_selection}

To select isolated primary (IP) galaxies we use a method similar to that of \citet{kauffmann_etal13}, who selected a volume-limited sample of galaxies with $\logm>9.25$ and $0.017<z<0.03$ from SDSS.  They then defined ``central" galaxies of stellar mass \mstar as those in their sample with no other galaxies with stellar mass greater than $\mstar/2$ within a projected radius of 500~kpc and with a velocity difference less than 500~\kms.

Here we use the following selection criteria to identify IP galaxies in PRIMUS:~IP \emph{candidates} must have at least twice the stellar mass of the mass completeness limit \mlim \citep[see][]{moustakas_etal13} for their redshift and galaxy type (star-forming or quiescent): ${\mstar \ge \mlim(z,\,{\rm galaxy\,type}) - \log(\mfrac)}$ where ${\mfrac=0.5}$.

We then define IP \emph{galaxies} as IP candidates with no neighbors with ${\mstar > \mfrac\,\mIP}$ within a projected distance of 500 comoving kpc (ckpc) and $\pm 2\sigma_{z}$ in redshift space, where $\sigma_{z}$ is defined by the correspondence between PRIMUS redshifts and higher-resolution spectroscopic redshifts in PRIMUS fields \citep[see][for details]{coil_etal11}.

These isolation criteria are chosen because they yield IP samples that balance the competing needs of a large sample size and high completeness (the fraction of true central galaxies correctly identified as isolated), while minimizing sample contamination (the fraction of galaxies identified as isolated but which are actually satellite galaxies). Below in section \S\ref{sec:results_comp_contam} we estimate the completeness and contamination of PRIMUS IP samples as a function of redshift and stellar mass for different values of \rp, \mfrac, and multiples of \sigmaz by applying the same isolation criteria to mock catalogs designed to match PRIMUS data.

It is possible for galaxies near the edge of the survey area to be incorrectly classified as isolated if in reality they have a sufficiently massive neighbor that would be within a projected distance of 500~ckpc that happens to lie outside the survey area. \citet{berti_etal17} investigated the potential for this effect to contaminate IP samples in PRIMUS by visually inspecting the distribution of IPs near the survey edges. They found that the spatial density of IPs does not rise substantially near the survey edges and concluded that false detections near edges do not significantly impact IP selection.

\subsection{Galaxy Samples}\label{sec:gal_samples}

A main goal of this work is investigate the dependence of clustering amplitude on both stellar mass and sSFR for IP galaxies (which we use a proxy for central galaxies) at $z\sim0.5$ in PRIMUS.
After classifying all PRIMUS galaxies in the full sample as star-forming or quiescent as described in \S\ref{sec:sfq}, and then selecting IP galaxies from the full sample, we next divide IP galaxies into various samples for which we measure clustering amplitude and absolute bias on two-halo scales (see \S\ref{sec:methods}).

The ``star-forming/quiescent split" run divides IP galaxies into star-forming and quiescent samples in two redshift bins to compare the clustering of star-forming versus quiescent IP galaxies at fixed stellar mass. The "main sequence split" run further divides IP galaxies within the star-forming and quiescent main sequences into samples above and below each main sequences. This allows us to test the dependence of clustering amplitude on sSFR at fixed stellar mass for IP galaxies \emph{within} the star-forming and quiescent main sequences.

\citet{coil_etal17} make similar measurements for all PRIMUS galaxies, and the IP galaxy samples we use here are analogous to their Run 1 (``star-forming/quiescent split") and Run 2 (``main sequence split") samples.
Because here we are concerned with IP galaxies and the density of identifiable IP galaxies in PRIMUS decreases significantly with redshift, the maximum redshift of any of our samples is $z=0.9$. Histograms of the redshift and stellar mass distributions of all IP galaxy samples are shown in Figures~\ref{fig:primus_redshift_hist} and \ref{fig:primus_mass_hist}, respectively. Table~\ref{tab:gal_samples} summarizes the properties of all galaxy samples.

\subsubsection{Star-forming/Quiescent Split}\label{sec:sfq_split}

We first define samples of star-forming and quiescent IP galaxies above the PRIMUS mass completeness limits in ``low" (${0.2 < z < 0.5}$) and ``high" (${0.5 < z < 0.9}$) redshift bins. Galaxies are classified as star-forming or quiescent by the same criteria used in \citet{coil_etal17} and \citet{berti_etal17}:~a galaxy with redshift $z$ is star-forming (quiescent) if it lies above (below) Equation~\ref{eq:sfr_cut} in the stellar mass-sSFR plane.

While our star-forming and quiescent IP galaxy populations are statistically complete (after applying the targeting and completeness weights described above), and while we require IP galaxies to be above the stellar mass completeness limits of the survey, the median stellar masses and redshifts of the star-forming and quiescent IP populations differ, due to differences in the stellar mass functions of star-forming and quiescent galaxies \citep[see][]{moustakas_etal13}.

The filled blue and red histograms in the top panel of Figure~\ref{fig:primus_redshift_hist} show the redshift distributions of {\em all} star-forming and quiescent PRIMUS IP galaxies, respectively. The unfilled histograms show stellar mass-matched and redshift-matched star-forming and quiescent IP galaxy samples, each of which is a subset of the corresponding filled histogram.

The top panels of Figure~\ref{fig:primus_mass_hist} show the stellar mass distributions of all star-forming and quiescent PRIMUS galaxies in low ($0.2<z<0.5$) and high ($0.5<z<0.9$) redshift bins, all star-forming and quiescent IP galaxies in each redshift bin, and stellar mass-matched samples of star-forming and quiescent IP galaxies within each bin. For example, the low redshift star-forming and quiescent IP galaxy samples have median stellar masses of $10^{10.3}~\msun$ and $10^{10.7}~\msun$, respectively, while the stellar mass-matched IP galaxy samples both have a median stellar mass of $\sim10^{10.6}~\msun$. We obtain stellar mass-matched IP galaxy samples by selecting all galaxies within the intersection of the stellar mass histograms of all star-forming IP galaxies and all quiescent IP galaxies. For the rest of this paper we use the stellar mass-matched IP galaxy samples for the ``star-forming/quiescent split" run (hatched histograms in the top panels of Figure~\ref{fig:primus_mass_hist} and outlined red and blue histograms in the top panel of Figure~\ref{fig:primus_redshift_hist}).

\subsubsection{Main Sequence Split}\label{sec:ms_split}

To investigate the clustering amplitude dependence of IP galaxies \emph{within} the star-forming and quiescent main sequences we divide the star-forming and quiescent IP galaxy populations in the redshift range ${0.2 < z < 0.7}$ each into two samples of approximately equal size and with the same median redshift (see the lower panels of Fig.~\ref{fig:primus_redshift_hist}) and stellar mass (see the lower panels of Fig.~\ref{fig:primus_mass_hist}).
Because here we split the IP galaxy population into four samples we use a wider redshift range than either used for the star-forming/quiescent split IP galaxy samples to maintain sufficiently large sample sizes and reduce uncertainty.

We split the star-forming IP galaxy population into ``star-forming above" (SF-above) and ``star-forming below" (SF-below) samples based on whether each galaxy is above or below the redshift-dependent star-forming main sequence (Eq.~\ref{eq:primus_ms_cut_blue}; solid blue lines in Fig.~\ref{fig:primus_ms_cuts}) evaluated at the redshift of that galaxy.
The quiescent population is similarly divided into ``quiescent above" (Q-above) and ``quiescent below" (Q-below) based on whether each galaxy falls above or below the redshift-dependent quiescent main sequence (Eq.~\ref{eq:primus_ms_cut_red}; solid red lines in Fig.~\ref{fig:primus_ms_cuts}). 

To identify the star-forming and quiescent main sequences as a function of redshift we first divide the star-forming and quiescent populations into redshift bins of width $\Delta z=0.1$ from $z=0.2$ to $z=0.7$. Within each redshift bin we divide the star-forming and quiescent populations separately into narrow bins in stellar mass, and then find the median sSFR separately for star-forming and quiescent galaxies within each stellar mass bin. For each galaxy type (star-forming or quiescent) in each redshift bin we fit a linear relation between median sSFR and mean stellar mass:
\begin{subequations}
  \begin{align}
	\log({\rm sSFR})^{\rm SF}_{\rm med} &= a_0^{\rm SF}(z)\log(M_*/M_{\odot}) + a_1^{\rm SF}(z)\label{eq:masssfr_coeffs00}\\
	\log({\rm sSFR})^{\rm Q}_{\rm med} &= a_0^{\rm Q}(z)\log(M_*/M_{\odot}) + a_1^{\rm Q}(z)\label{eq:masssfr_coeffs01}.
  \end{align}
\label{eq:masssfr_coeffs}
\end{subequations}

To account for the redshift dependence of the coefficients $a_i^{\rm SF}(z)$ and $a_i^{\rm Q}(z)$ in Equation~\ref{eq:masssfr_coeffs} (where $i=0,1$) we fit a second set of linear relations between the mean redshift in each bin and the fitted values of $a_i^{\rm SF}(z)$ and $a_i^{\rm Q}(z)$ at each mean redshift:
\begin{subequations}
  \begin{align}
	a_0^{\rm SF}(z) &= b_0^{\rm SF}z + c_0^{\rm SF} &a_1^{\rm SF}(z) &= b_1^{\rm SF}z + c_1^{\rm SF}\label{eq:z_coeffs00} \\
	a_0^{\rm Q}(z) &= b_0^{\rm Q}z + c_0^{\rm Q} &a_1^{\rm Q}(z) &= b_1^{\rm Q}z + c_1^{\rm Q}\label{eq:z_coeffs01}
  \end{align}
\label{eq:z_coeffs}
\end{subequations}

Combining Eqs.~\ref{eq:masssfr_coeffs00} and \ref{eq:masssfr_coeffs01} with Eqs.~\ref{eq:z_coeffs00} and \ref{eq:z_coeffs01} gives the final redshift-dependent equations for the star-forming and quiescent main sequences. Star-forming IP galaxies are classified as above or below the star-forming main sequence based on whether they fall above or below the following cut in the stellar mass-sSFR plane:
\begin{align}
  	\log({\rm sSFR})^{\rm SF}_{\rm MS}
  	&= \left(b_0^{\rm SF}z + c_0^{\rm SF}\right)\log\left[\frac{M_*}{M_{\odot}}\right] + b_1^{\rm SF}z + c_1^{\rm SF} \nonumber \\
	&= 2.02z - (0.08z + 0.67)\log\left[\frac{M_*}{M_{\odot}}\right] - 3.53.
\label{eq:primus_ms_cut_blue}
\end{align}

Similarly, the cut in the stellar mass-sSFR plane that divides quiescent IP galaxies into samples above and below the quiescent main sequence is given by
\begin{align}
	\log({\rm sSFR})^{\rm Q}_{\rm MS}
	&= \left(b_0^{\rm Q}z + c_0^{\rm	Q}\right)\log\left[\frac{M_*}{M_{\odot}}\right] + b_1^{\rm Q}z + c_1^{\rm Q} \nonumber \\
	&= 5.4z - (0.3z + 0.38)\log\left[\frac{M_*}{M_{\odot}}\right] - 8.64.
\label{eq:primus_ms_cut_red}
\end{align}

Figure~\ref{fig:primus_ms_cuts} shows stellar mass versus sSFR for the four main sequence split IP galaxy samples (SF-above, SF-below, Q-above, Q-below) in three redshift bins:~${0.2 < z < 0.35}$, ${0.35 < z < 0.5}$, and ${0.5 < z < 0.7}$, and the star-forming and quiescent main sequence cuts (Eqs.~\ref{eq:primus_ms_cut_blue} (solid blue lines) and \ref{eq:primus_ms_cut_red} (solid red lines), respectively) evaluated at the mean redshift of each bin.

\input{gal_samples.tex}

%% file: gal_samples.tex
\begin{deluxetable*}{llrrrrrrrrrr}[tb]
\tablecaption{PRIMUS IP galaxy samples.
\label{tab:gal_samples}
}
\tablewidth{0pt}

\tablehead{
\colhead{\multirow{2}{*}{Run}} &
\colhead{\multirow{2}{*}{Sample Name}} &
\colhead{\multirow{2}{*}{$N_{\rm gal}$}} &
\multicolumn{3}{c}{Redshift} &
\multicolumn{3}{c}{$\log(M_*/M_{\odot})$} &
\multicolumn{3}{c}{$\log($sSFR/yr$^{-1})$} \\
{} & {} & {} &
\colhead{min} & \colhead{mean} & \colhead{max} &
\colhead{min} & \colhead{mean} & \colhead{max} &
\colhead{min} & \colhead{mean} & \colhead{max}
}
\startdata
Star-forming/quiescent split & Q-lowz & 2892 & 0.2 & 0.36 & 0.5 & 9.47 & 10.63 & 11.61 & -13.07 & -11.86 & -10.89 \\
& SF-lowz & 2706 & 0.2 & 0.38 & 0.5 & 9.38 & 10.63 & 11.55 & -11.46 & -10.24 & -8.06 \\
& Q-highz & 3179 & 0.5 & 0.66 & 0.9 & 10.43 & 10.98 & 11.88 & -12.68 & -11.50 & -10.63 \\
& SF-highz & 3331 & 0.5 & 0.71 & 0.9 & 10.38 & 10.98 & 11.99 & -11.13 & -10.17 & -8.49 \\
\\
Main sequence split & Q-below & 3106 & 0.2 & 0.47 & 0.7 & 9.47 & 10.79 & 11.75 & -13.37 & -11.98 & -11.23 \\
& Q-above & 2447 & 0.2 & 0.47 & 0.7 & 9.56 & 10.82 & 11.76 & -12.40 & -11.46 & -10.83 \\
& SF-below & 6598 & 0.2 & 0.47 & 0.7 & 8.98 & 10.36 & 11.67 & -11.46 & -10.28 & -9.30 \\
& SF-above & 5486 & 0.2 & 0.46 & 0.7 & 8.99 & 10.35 & 11.70 & -10.82 & -9.63 & -7.94 \\
\enddata
\end{deluxetable*}


%% file: mocks.tex
\section{Mock Galaxy Catalogs}\label{sec:mocks}

\begin{figure*}[bt]
\centering
\includegraphics[width=0.8\linewidth]{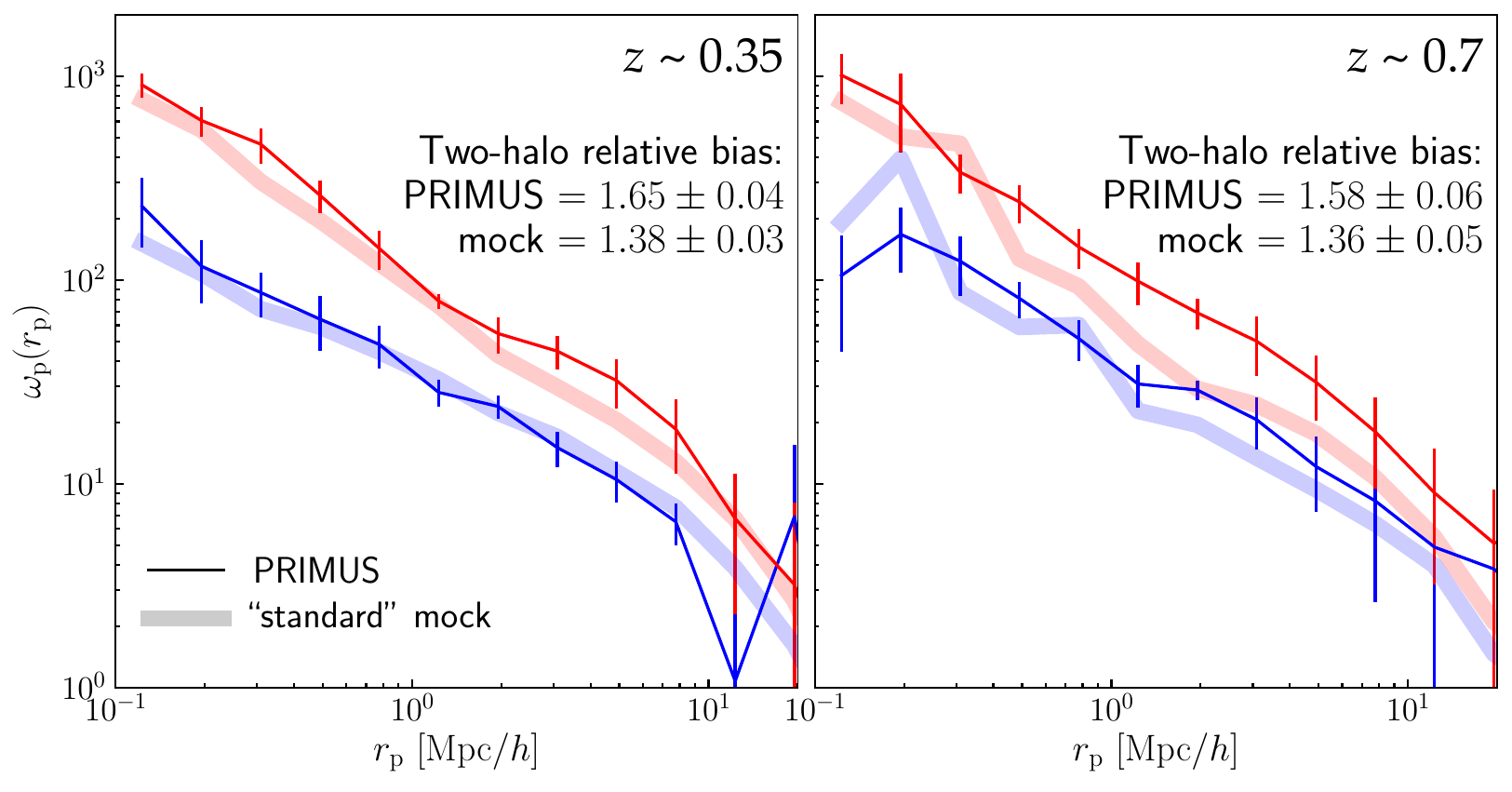}
\caption{
Clustering amplitude, \wprp, for stellar mass-matched samples of all star-forming (thin blue lines) and all quiescent (thin red lines) PRIMUS galaxies at $0.2 < z < 0.5$ (left) and $0.5 < z < 0.9$ (right). Also shown is \wprp for all star-forming (thick blue lines) and quiescent (thick red lines) mock galaxy samples from the PRIMUS-matched ``standard" mock galaxy catalogs at $z=0.35$ (left) and $z=0.7$ (right) in the same stellar mass ranges as PRIMUS samples.
The clustering of all galaxies in the PRIMUS-matched mock catalog agrees with PRIMUS at low redshift. At high redshift galaxies in the PRIMUS-matched mock catalog are less clustered than PRIMUS galaxies, but the \emph{relative} bias of quiescent versus star-forming galaxies at fixed stellar mass agrees with PRIMUS within uncertainty.
}
\label{fig:wp_all_gal}
\end{figure*}

We create mock galaxy samples to help interpret our observational results. The mock galaxy catalogs we use are based on UniverseMachine \citep{behroozi_etal18}. By empirically modeling the dependence of galaxy SFR as a function of halo mass, halo accretion rate, and redshift, the UniverseMachine makes predictions for the star formation history of galaxies across time, connecting these histories to the assembly history of dark matter halos. Predicting star formation history by modeling galaxy--halo co-evolution has a long history in the field of semi-analytic modeling \citep[e.g.,][]{kauffmann_etal93, somerville_primack99, benson12, somerville_dave15}. Recent progress in empirically modeling galaxy formation has enabled a new generation of data-driven models to self-consistently predict galaxy assembly histories across cosmic time \citep{becker15,cohn17,moster_etal17}. As shown in \citet{behroozi_etal18}, the UniverseMachine model has sufficient flexibility to capture a wide range of statistics summarizing the observed galaxy distribution across redshift, including stellar mass functions, quiescent fractions, and two-point galaxy clustering. 

We use the publicly available UniverseMachine code to generate synthetic galaxies populating snapshots in the MDPL2 simulation \citep{klypin_etal16}.\footnote{In practice we use a $250\times250\times400~{\rm Mpc}/h$ cutout of MDPL2.}
The snapshots we use are at the mean redshifts of our low and high redshift star-forming/quiescent split PRIMUS IP galaxy samples:~$z=0.35$ and $z=0.7$, respectively.

In each snapshot, every subhalo identified by Rockstar in MDPL2 is populated with a synthetic galaxy.\footnote{UniverseMachine mock catalogs include an additional treatment of orphan galaxies that reside in subhalos no longer identified by Rockstar; see Appendix D of \citet{behroozi_etal18} for further details.} 
Briefly, star-formation histories of UniverseMachine galaxies are modeled as follows. First, at each redshift, the distribution of SFRs of synthetic galaxies is modeled with a double Gaussian, one for quenched galaxies and one for star-forming galaxies. The locations and relative heights of the two SFR peaks have a parameterized dependence on both $V_{\rm max}$ (the maximum circular halo velocity) and redshift. At fixed $V_{\rm max}$ UniverseMachine allows for the possibility that galaxy SFR is correlated with (sub)halo $\Delta V_{\rm max}\vert_{\tau_{\rm dyn}}$, the change in $V_{\rm max}$ over the last dynamical time, ${\tau_{\rm dyn} \equiv (\frac{4}{3}\pi G \rho_{\rm vir})^{-1/2}}$, where $\rho_{\rm vir}$ is the virial overdensity. Conditional Abundance Matching (CAM) is used to implement the correlation between SFR and $\Delta V_{\rm max}\vert_{\tau_{\rm dyn}}$, while the strength of this correlation is parameterized to allow for possible dependence upon both halo mass and redshift.
We then tailor these mock galaxy catalogs to mimic PRIMUS galaxy targeting with the process described in \S\ref{sec:mock_match} below.
Throughout this paper we refer to these mock galaxy catalogs (after matching them to observed PRIMUS data) as our ``standard" mocks.

Figure~\ref{fig:wp_all_gal} shows \wprp separately for star-forming and quiescent PRIMUS galaxies at fixed stellar mass in low ($0.2<z<0.5$) and high ($0.5<z<0.9$) redshift bins. Also shown for comparison is \wprp for the corresponding mock galaxy samples from the $z=0.35$ and $z=0.7$ ``standard" mock galaxy catalogs. At low redshift ($z\sim0.35$) the relative bias on two-halo scales of all quiescent to all star-forming PRIMUS galaxies at fixed stellar mass is $1.65\pm0.04$, while in the ``standard" mock catalog the relative bias is $1.38\pm0.03$. At high redshift ($z\sim0.7$) this two-halo relative bias is $1.58\pm0.06$ in PRIMUS and $1.36\pm0.05$ in the ``standard" mock catalog.  While the relative biases in the PRIMUS galaxy samples and the mock galaxy samples are not identical, they are similar, and the clustering properties of the mock catalogs are sufficient for the purposes used here.

In this work we use IP galaxies in PRIMUS as an observational proxy for central galaxies.
Our investigation will benefit from mock galaxy catalogs with a strong intrinsic signal\footnote{By ``signal" we mean specifically a difference in clustering amplitude between quiescent and star-forming galaxies at fixed stellar mass.} so that we can conduct a systematic study of how various measurement choices weaken or change the underlying clustering difference between quiescent and star-forming central galaxies. To facilitate such testing, we create an alternate version of the best-fit UniverseMachine model and an associated ``modified" mock galaxy catalog at $z=0.35$, in which we enhance the strength of the correlation between central galaxy SFR and host halo accretion rate ($dM_{\rm halo}/dt$). The observable effect of this enhancement is a correlation between galaxy sSFR and clustering amplitude at fixed stellar mass for galaxies \emph{within} the quiescent main sequence (and to a lesser extent within the star-forming main sequence), consistent with the observations of \citet{coil_etal17} for PRIMUS galaxies. The relative bias between star-forming and quiescent galaxies is unchanged compared to the ``standard" $z=0.35$ mock galaxy catalog.

We use CAM to model this enhanced correlation \citep[see][]{hearin_etal14, watson_etal15}.
At fixed stellar mass, and separately for star-forming and quiescent centrals, we assume that sSFR varies monotonically with halo accretion rate, with stochasticity such that sSFR and $dM_{\rm halo}/dt$ exhibit a $50\%$  rank-order correlation coefficient.
The bin-free implementation of CAM in Halotools \citep{hearin_etal17} exactly preserves ${\rm P(SFR\vert M_*)}$, and so the one-point correlation functions in our mock are in exact statistical agreement with the best-fit UniverseMachine model. The two-halo relative bias of quiescent versus star-forming galaxies at fixed stellar mass in the ``modified" mock is $1.38\pm0.03$, in precise agreement with the ``standard" $z=0.35$ mock catalog.

\subsection{Matching Mock Catalogs to PRIMUS}\label{sec:mock_match}

To compare the clustering of PRIMUS IP galaxies to that of IP and central galaxies in mock catalogs we require mock catalogs that match the PRIMUS dataset at a given redshift as closely as possible. Specifically, mock catalogs should have the same stellar mass and sSFR distributions, galaxy number density, and line-of-sight position (related to redshift) uncertainty as PRIMUS data.

We use the following process to match our mock catalogs to the PRIMUS survey dataset:
\begin{enumerate}
\item \textbf{Match the joint stellar mass and sSFR distribution.} For each mock catalog we create a normalized two-dimensional histogram of the joint stellar mass and sSFR distribution of all $z_{\rm quality} \ge 3$ PRIMUS galaxies within a given redshift range using bins of width 0.1 dex in stellar mass and 0.1 dex in sSFR. The redshift ranges used are $0.2<z_{\rm PRIMUS}<0.5$ and $0.5<z_{\rm PRIMUS}<0.9$ for the $z=0.35$ and $z=0.7$ ``standard" mock galaxy catalogs, respectively. For the $z=0.35$ ``modified" mock catalog we use $0.2<z_{\rm PRIMUS}<0.7$, the redshift range of the main sequence split PRIMUS IP galaxy samples.

We then randomly select from each bin the number of mock galaxies equal to the weight of that bin multiplied by the total number of mock galaxies in all bins with nonzero weight. This process eliminates mock galaxies with sufficiently low stellar mass and/or sSFR that they would not be PRIMUS targets and ensures that the remaining mock galaxies have the same stellar mass and sSFR distributions as PRIMUS galaxies within a given redshift range around the redshift of the mock.

\item \textbf{Match the mean galaxy number density.} We then randomly down-sample the mock catalog so that it has the same number density as the mean observed PRIMUS number density over the relevant redshift range:~$0.3<z_{\rm PRIMUS}<0.4$ for $\zmock=0.35$ and $0.65<z_{\rm PRIMUS}<0.75$ for $\zmock=0.7$.
We use the mean density of the two largest PRIMUS fields:~CDFS and XMM. In \S\ref{sec:density_test} below we test the effect of using higher and lower densities on IP galaxy selection and the relative bias of IP and central galaxy samples in the mock catalogs.

\item \textbf{Add PRIMUS-like redshift-space distortions.} While mock galaxy catalogs contain precise spatial information in three dimensions, PRIMUS redshifts have an uncertainty of $\sigmaz \simeq0.005(1+z)$, which translates to an uncertainty in galaxy position in the comoving line-of-sight distance $r_z$. To emulate this uncertainty in the mock catalogs we add $\Delta r_z$ (Mpc$/h$) to each mock galaxy's $r_z$ coordinate such that ${r^{\rm noisy}_z = r_z + \Delta r_z}$, where $\Delta r_z$ is randomly drawn from a normal distribution of width $\sigma_{r_z}(z)$ (Mpc$/h$). We compute $\sigma_{r_z}(z)$ by using the relationship between peculiar velocity $v_{\rm phys}$ and redshift to express the PRIMUS peculiar velocity uncertainty $\sigma_{v_{\rm phys}}$ in terms of redshift:
\begin{align}
z \simeq v_{\rm phys}/c \rightarrow \sigma_{v_{\rm phys}} &= c\,\sigmaz \nonumber\\
&= c\times0.005(1+z).
\end{align}
Physical distance $r_{z,\,{\rm phys}}$ is related to comoving line-of-sight distance $r_z$ by the scale factor $a(z)$:
\begin{equation}
r_{z,\,{\rm phys}} = a(z)\,r_z,
\end{equation}
and to peculiar velocity $v_{\rm phys}$ by $H(z)$:
\begin{equation}
r_{z,\,{\rm phys}} = v_{\rm phys}/H(z).
\end{equation}
Combining the three equations above we have
\begin{align}
\sigma_{r_z}(z) &= \frac{\sigma_{v_{\rm phys}}}{a(z)\,H(z)} \nonumber\\
&= \frac{c\times0.005(1+z)^2}{H_0\sqrt{\Omega_{\rm M}(1+z)^3 + \Omega_k(1+z)^2 + \Omega_{\Lambda}}}.
\end{align}

For example, $\sigma_{r_z}\simeq22.8$ Mpc$/h$ at $z=0.35$ and $\sigma_{r_z}\simeq29.4$ Mpc$/h$ at $z=0.7$.

\end{enumerate}

In \S\ref{sec:density_test} below we compare IP galaxy selection and clustering results in the mock catalogs with and without added PRIMUS redshift-space uncertainty.

\input{mock_samples.tex}

\subsection{IP Galaxy Selection and Mock Samples}\label{sec:mock_samples}

IP galaxies were selected in the PRIMUS-matched ``standard" and ``modified" mock galaxy catalogs using the same criteria used to select IP galaxies in PRIMUS, where a cylinder depth of $\pm2\sigma_{r_z}(\zmock)$~Mpc/$h$ was used instead of $\pm2\sigmaz$ in redshift space because each mock is a snapshot at a single redshift.

\begin{figure}[bt]
\centering
\includegraphics[width=\linewidth]{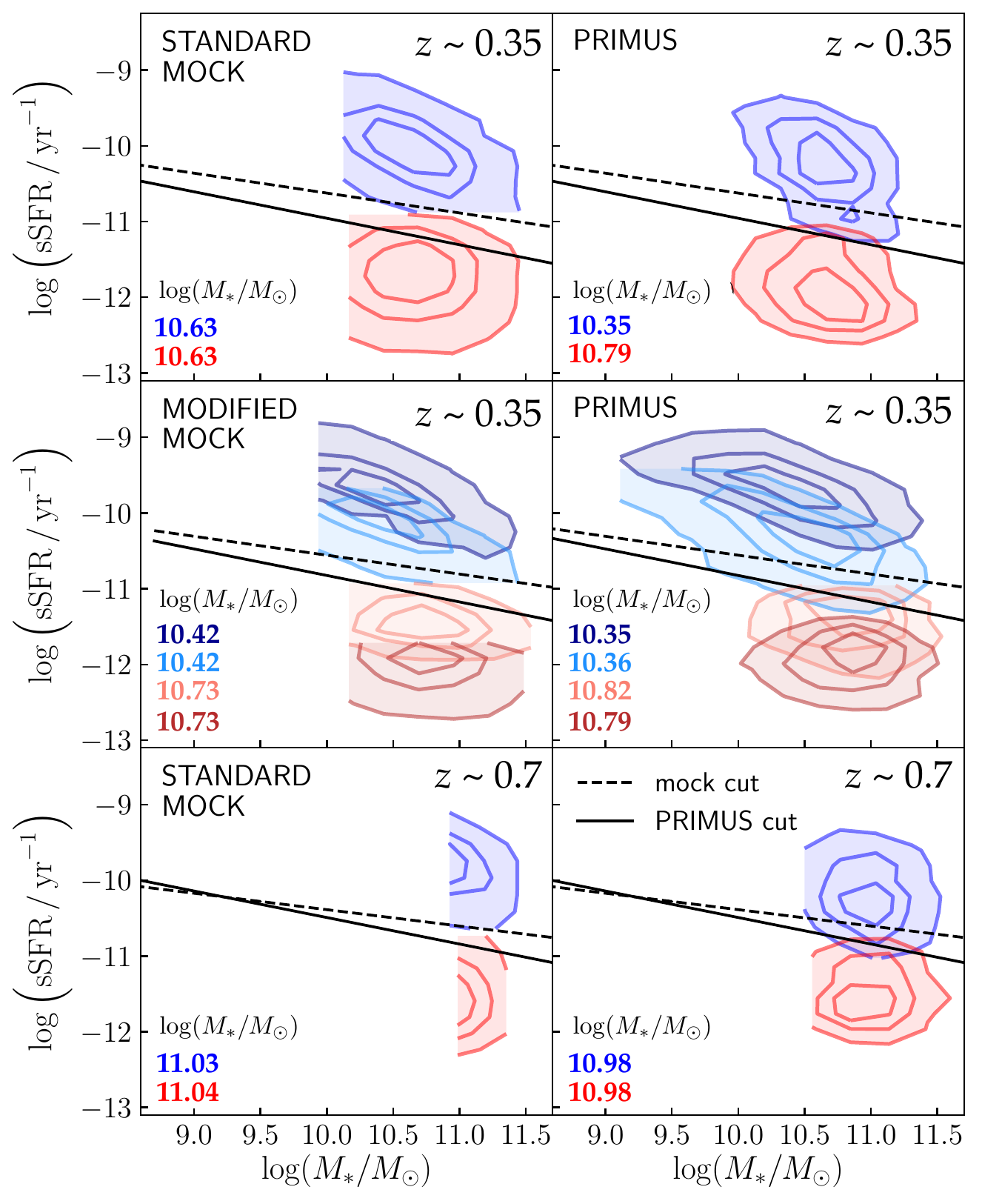}
\caption{Stellar mass versus sSFR for the low redshift (top row) and high redshift (bottom row) ``star-forming/quiescent split" IP galaxy samples and four ``main sequence split" IP galaxy samples (middle row) in the PRIMUS-matched mock galaxy catalogs (left column) and in PRIMUS (right column). Each panel lists the median stellar mass of each galaxy sample.
The PRIMUS galaxy type cut (Eq.~\ref{eq:sfr_cut}; solid black line) does not quite fit the minimum of the bimodal stellar mass-sSFR distribution of the corresponding PRIMUS-matched mock catalog. We therefore use a slightly different cut (Eq.~\ref{eq:mock_masssfr_cut}; dashed black line) to divide the PRIMUS-matched mock galaxy catalogs into star-forming and quiescent galaxy samples.
}
\label{fig:masssfr}
\end{figure}

IP galaxies from ``standard" mock catalogs were divided into star-forming and quiescent samples, while IP galaxies from the ``modified" mock catalog were divided into samples above and below the star-forming and quiescent main sequences. However, the same  cuts used to divide star-forming and quiescent galaxies in PRIMUS do not precisely align with the PRIMUS-matched mock catalogs. This is shown in Fig.~\ref{fig:masssfr} for the $z=0.35$ ``modified" mock catalog; solid lines show the PRIMUS cuts (Eq.~\ref{eq:sfr_cut}), while the dashed lines show the redshift-dependent cuts that are fit to the minimum of the bimodal mock catalog stellar mass-sSFR distributions, which are given by the following relation:
\begin{equation}
\log({\rm sSFR}) = -0.64z + (0.13z - 0.31)\log\left[\frac{M_*}{M_{\odot}}\right] - 7.77.
\label{eq:mock_masssfr_cut}
\end{equation}

To divide star-forming and quiescent IP galaxies in the ``modified" mock catalog into samples above and below the main sequence of each population we found the median stellar mass in each of a series of narrow mass bins for the star-forming (quiescent) mock IP galaxy population. The ``SF-above" (``SF-below") mock IP galaxy sample consists of all star-forming mock IP galaxies with stellar masses above (below) the median star-forming IP galaxy stellar mass in for their mass bin. Similarly, the ``Q-above" (``Q-below") mock IP galaxy sample is all quiescent mock IP galaxies with stellar masses above (below) the relevant median quiescent mock IP galaxy stellar mass.
This method splits the star-forming and quiescent mock IP galaxy populations each into two samples of equal size and mean stellar mass. Table~\ref{tab:mock_samples} summarizes all galaxy samples from the ``standard" and ``modified" mock galaxy catalogs.

%% file: mock_samples.tex
\begin{deluxetable*}{llrcrrrrrr}[tb]
\tablecaption{Mock galaxy samples.
\label{tab:mock_samples}
}
\tablewidth{0pt}

\tablehead{
\colhead{\multirow{2}{*}{Run}} &
\colhead{\multirow{2}{*}{Sample Name}} &
\colhead{\multirow{2}{*}{$N_{\rm gal}$}} &
\colhead{\multirow{2}{*}{$z_{\rm mock}$}} &
\multicolumn{3}{c}{$\log(M_*/M_{\odot})$} &
\multicolumn{3}{c}{$\log($sSFR/yr$^{-1})$} \\
{} & {} & {} & {} &
\colhead{min} & \colhead{mean} & \colhead{max} &
\colhead{min} & \colhead{mean} & \colhead{max}
}
\startdata
Star-forming/quiescent split
 & SF-lowz-IP & 23608 & 0.35 & 10.00 & 10.63 & 11.59 & -10.98 & -10.05 & -8.75 \\
(``standard" mock catalogs)
 & Q-lowz-IP & 23608 & 0.35 & 10.04 & 10.63 & 11.60 & -13.04 & -11.72 & -10.80 \\
 & SF-highz-IP & 4141 & 0.7 & 10.80 & 11.03 & 11.59 & -10.74 & -9.87 & -8.95 \\
 & Q-highz-IP & 4141 & 0.7 & 10.86 & 11.04 & 11.59 & -12.76 & -11.54 & -10.62 \\
\\
 & SF-lowz-cen & 26257 & 0.35 & 10.00 & 10.62 & 11.58 & -10.98 & -10.03 & -8.85 \\
 & Q-lowz-cen & 26257 & 0.35 & 10.04 & 10.62 & 11.60 & -13.02 & -11.69 & -10.79 \\
 & SF-highz-cen & 3843 & 0.7 & 10.80 & 11.02 & 11.59 & -10.74 & -9.86 & -8.72 \\
 & Q-highz-cen & 3843 & 0.7 & 10.86 & 11.03 & 11.58 & -12.76 & -11.51 & -10.62 \\
\\
 & SF-lowz-all & 42321 & 0.35 & 10.00 & 10.58 & 11.59 & -10.98 & -10.02 & -8.83 \\
 & Q-lowz-all & 42321 & 0.35 & 10.04 & 10.59 & 11.57 & -13.03 & -11.74 & -10.79 \\
 & SF-highz-all & 5018 & 0.7 & 10.80 & 11.02 & 11.59 & -10.74 & -9.86 & -8.64 \\
 & Q-highz-all & 5018 & 0.7 & 10.86 & 11.02 & 11.58 & -12.76 & -11.54 & -10.62 \\
 \\
Main sequence split
 & SF-above-IP & 22140 & 0.35 & 9.81 & 10.42 & 11.57 & -12.50 & -9.66 & -8.49 \\
(``modified" mock catalog)
 & SF-below-IP & 22140 & 0.35 & 9.81 & 10.42 & 11.57 & -12.83 & -10.13 & -8.44 \\
 & Q-above-IP & 11187 & 0.35 & 10.04 & 10.73 & 11.69 & -13.23 & -11.43 & -9.63 \\
 & Q-below-IP & 11187 & 0.35 & 10.04 & 10.73 & 11.62 & -13.33 & -12.03 & -10.08 \\
\\
 & SF-above-cen & 29562 & 0.35 & 9.81 & 10.40 & 11.57 & -12.50 & -9.65 & -8.41 \\
 & SF-below-cen & 29562 & 0.35 & 9.81 & 10.40 & 11.57 & -12.82 & -10.11 & -8.41 \\
 & Q-above-cen & 11989 & 0.35 & 10.04 & 10.70 & 11.69 & -13.23 & -11.41 & -9.40 \\
 & Q-below-cen & 11989 & 0.35 & 10.04 & 10.70 & 11.65 & -13.33 & -12.00 & -10.08 \\
\\
 & SF-above-all & 38119 & 0.35 & 9.81 & 10.39 & 11.57 & -12.50 & -9.65 & -8.54 \\
 & SF-below-all & 38119 & 0.35 & 9.81 & 10.39 & 11.57 & -12.83 & -10.12 & -8.48 \\
 & Q-above-all & 19834 & 0.35 & 10.04 & 10.68 & 11.69 & -13.23 & -11.45 & -9.40 \\
 & Q-below-all & 19834 & 0.35 & 10.04 & 10.68 & 11.66 & -13.33 & -12.06 & -10.08 \\

\enddata
\end{deluxetable*}

%% file: methods.tex
\section{Methods}\label{sec:methods}

\subsection{Clustering Measurements}\label{sec:clust}

To quantify the clustering of IP galaxies we measure the two-point correlation function $\xir$, which determines the excess probability above a random Poisson distribution of a pair of galaxies having a given physical separation.

Following the methods of \citet{coil_etal17}, we first measure the cross-correlation function (CCF) of a given IP galaxy sample with a ``tracer" galaxy sample consisting of all PRIMUS galaxies with robust redshifts within the relevant redshift range. For CCF measurements we use the \citet{davis_etal83} estimator:~$\xirppi = GT/GR - 1$, where $GT$ is the sum of the weighted pair counts of IP galaxies ($G$) and tracer galaxies ($T$) and $GR$ is the sum of the weighted pair counts of IP galaxies and a random catalog that has the same projected spatial distribution and redshift distribution as the tracer sample. 
The random catalogs also account for the PRIMUS redshift success fraction discussed in \S\ref{sec:PRIMUS} above. Weights account for target selection in PRIMUS (see \S\ref{sec:targ_weight}) and allow us to create a statistically complete sample not subject to spatial biases.
Pair counts are measured in bins of projected distance \rp and line-of-sight distance $\pi$.

We also measure the auto-correlation function (ACF) of the tracer sample in the same volume as each IP galaxy sample using the \citet{landy_szalay93} estimator:
$\xirppi=(GG - 2GR)/RR + 1$,
where $GG$, $GR$, and $RR$ are weighted pair counts of galaxies in the galaxy-galaxy, galaxy-random, and random-random catalogs.
Weights again are used to account for the  PRIMUS target selection.

To account for redshift space distortions, we then obtain the projected correlation function \wprp by integrating all ACFs and CCFs along the line-of-sight to a maximum separation of $\pimax=40$~Mpc$/h$, chosen because the signal-to-noise ratio of \xirppi declines quickly for larger values of $\pi$.

Finally, we infer the ACF for a particular IP galaxy sample from the corresponding projected galaxy--tracer CCF, $\omega_{\rm GT}(\rp)$, and projected tracer ACF, $\omega_{\rm TT}(\rp)$:
\begin{equation}
\omega_{\rm GG}(\rp) = \frac{\omega^2_{\rm GT}(\rp)}{\omega_{\rm TT}(\rp)}.
\end{equation}
This method assumes that the spatial distributions of the IP and tracer galaxies in each sample are linearly related to the spatial distribution of the underlying dark matter, and that IP and tracer galaxies are well-mixed within dark matter halos. \citet{coil_etal17} test this assumption by comparing the direct ACF of both star-forming and quiescent PRIMUS galaxies with the ACF inferred from the galaxy--tracer CCF and find excellent agreement on both small and large projected scales.

\subsection{Relative Bias}\label{sec:bias_calc}

In this paper we focus on relative bias measurements to compare the clustering amplitudes of different galaxy samples at the same redshift.  
The relative bias of two galaxy samples is defined to be the square root of the ratio of their respective projected correlation functions and is a simple comparison of the clustering strengths of the two samples over a given length scale. We estimate the relative bias $b_{\rm rel}$ between two IP galaxy samples on the ``two-halo" scale of $1<\rp<10$~Mpc$/h$ by taking the mean of $\sqrt{\omega_1(\rp)/\omega_2(\rp)}$ over the scales of interest, where $\omega_1$ and $\omega_2$ are the values of the projected auto-correlation functions of the two galaxy samples at each \rp bin value between 1 and 10 Mpc$/h$.

\subsection{Error Estimation}\label{sec:error}

We estimate the uncertainty on \wprp using the jackknife resampling procedure described in \citet{coil_etal17}, which accounts for uncertainty due to cosmic variance. We use 10 jackknife samples across the four PRIMUS fields. The two larger fields, CDFS and XMM, are each divided along lines of constant RA and Dec into four subfields of roughly equal area on the sky. The smaller fields, COSMOS and ES1, are not subdivided as they are each approximately one fourth the size of the larger fields.

We calculate the projected correlation function \wprp for each jackknife sample $j$. The variance of \wprp is then
\begin{equation}
\sigma^2_{\wp}(\rp) = \frac{N-1}{N}\sum_j^N (\wprp - \omega_j(\rp))^2,
\end{equation}
\noindent where $N$ is the total number of jackknife samples and $\omega_j(\rp)$ is the projected correlation function for a given jackknife sample.  These errors are shown in figures of \wprp for individual bins in \rp. 

To estimate the uncertainty on the relative bias between two galaxy samples, we calculate $b_{\rm rel}$ for {\it each} jackknife sample $j$ and measure the variance of the relative bias across the samples: 
\begin{equation}
\sigma^2_{b_{\rm rel}} = \frac{N-1}{N}\sum_j^N (b_{\rm rel} - b_{{\rm rel},j})^2,
\end{equation}
\noindent where $b_{{\rm rel},j}$ is the relative bias for a given jackknife sample.  We therefore do not use the error bars on \wprp for each sample, but rather calculate the variance of the relative bias itself.

For the mock catalogs, to estimate the uncertainty on the two-halo scale relative bias for mock galaxy samples we repeat the pipeline described in \S\ref{sec:mock_match} 10 times to create 10 versions of the relevant PRIMUS-matched mock galaxy catalog, each with its own set of the relevant galaxy samples in Table~\ref{tab:mock_samples}. Then for a given sample or samples from each of the 10 versions of the mock galaxy catalog we calculate the relative bias, take the mean of all 10 values, and take the standard deviation of all 10 values as the error on the mean.

%% file: results_clustering.tex
\section{Isolated Primary Clustering Results}\label{sec:results_clustering}

\begin{figure*}[t]
\centering
\includegraphics[width=0.8\linewidth]{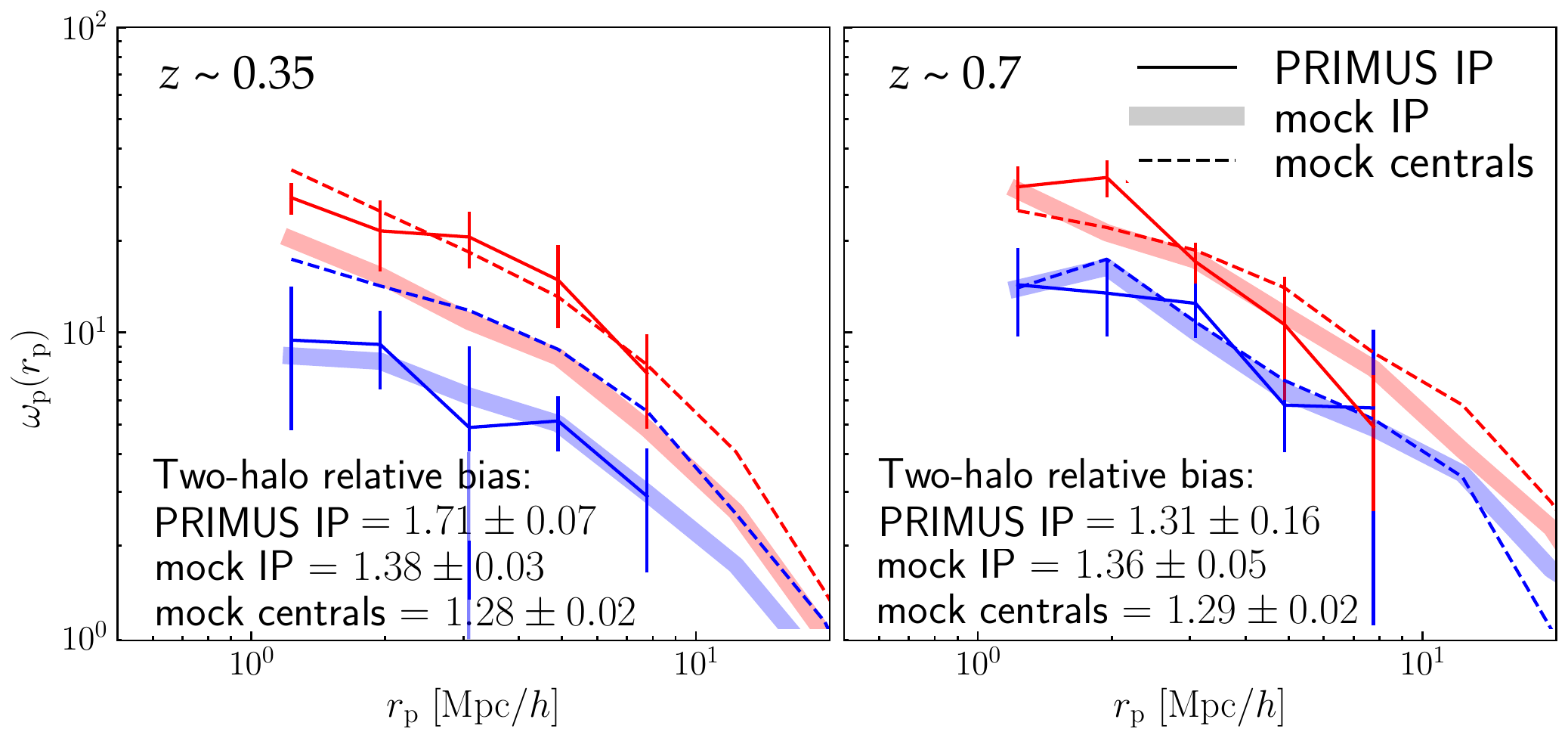}
\caption{
Clustering amplitude, \wprp, for star-forming (thin blue lines) and quiescent (thin red lines) PRIMUS IP galaxy samples at low ($0.2 < z < 0.5$; left) and high ($0.5 < z < 0.9$; right). Also shown are star-forming and quiescent IP galaxies (dashed blue and red lines) and central galaxies (thick blue and red lines) in low ($z=0.35$) and high ($z=0.7$) redshift PRIMUS-matched ``standard" mock galaxy catalogs.
Within each redshift bin all PRIMUS and mock galaxy samples have the same stellar mass distribution:~the mean stellar mass is $\sim10^{10.6}~\msun$ at {$0.2<z<0.5$} and $\sim10^{11.0}~\msun$ at {$0.5<z<0.9$} (see Tables \ref{tab:gal_samples} and \ref{tab:mock_samples}). At fixed stellar mass quiescent IP galaxies are more strongly clustered on two-halo scales than star-forming IP galaxies.
}
\label{fig:wp_sfq}
\end{figure*}

In this section we present two-halo relative bias measurements for the various IP galaxy samples from PRIMUS, and compare these results to the relative bias measurements for IP and central galaxy samples from the mock galaxy catalogs.
We first present results for the ``star-forming/quiescent split" samples in \S\ref{sec:sfqsplit}, and use the mock galaxy catalogs to refine the relative biases we measure in PRIMUS data.
In \S\ref{sec:mssplit} we present results for the ``main sequence split" galaxy samples in PRIMUS and the mock catalogs. These galaxy samples are selected to investigate whether we detect any dependence of clustering amplitude on sSFR for IP galaxies separately within the star-forming and quiescent main sequences (``intra-sequence" relative bias).
Finally in \S\ref{sec:density_test} we explore how factors such as galaxy number density and the presence of PRIMUS redshift errors affect the selection of IP galaxies and the relative biases of IP and central galaxies we measure in the mock galaxy catalogs.

\subsection{Star-forming/quiescent split}\label{sec:sfqsplit}

\begin{figure*}[t]
\centering
\includegraphics[width=0.8\linewidth]{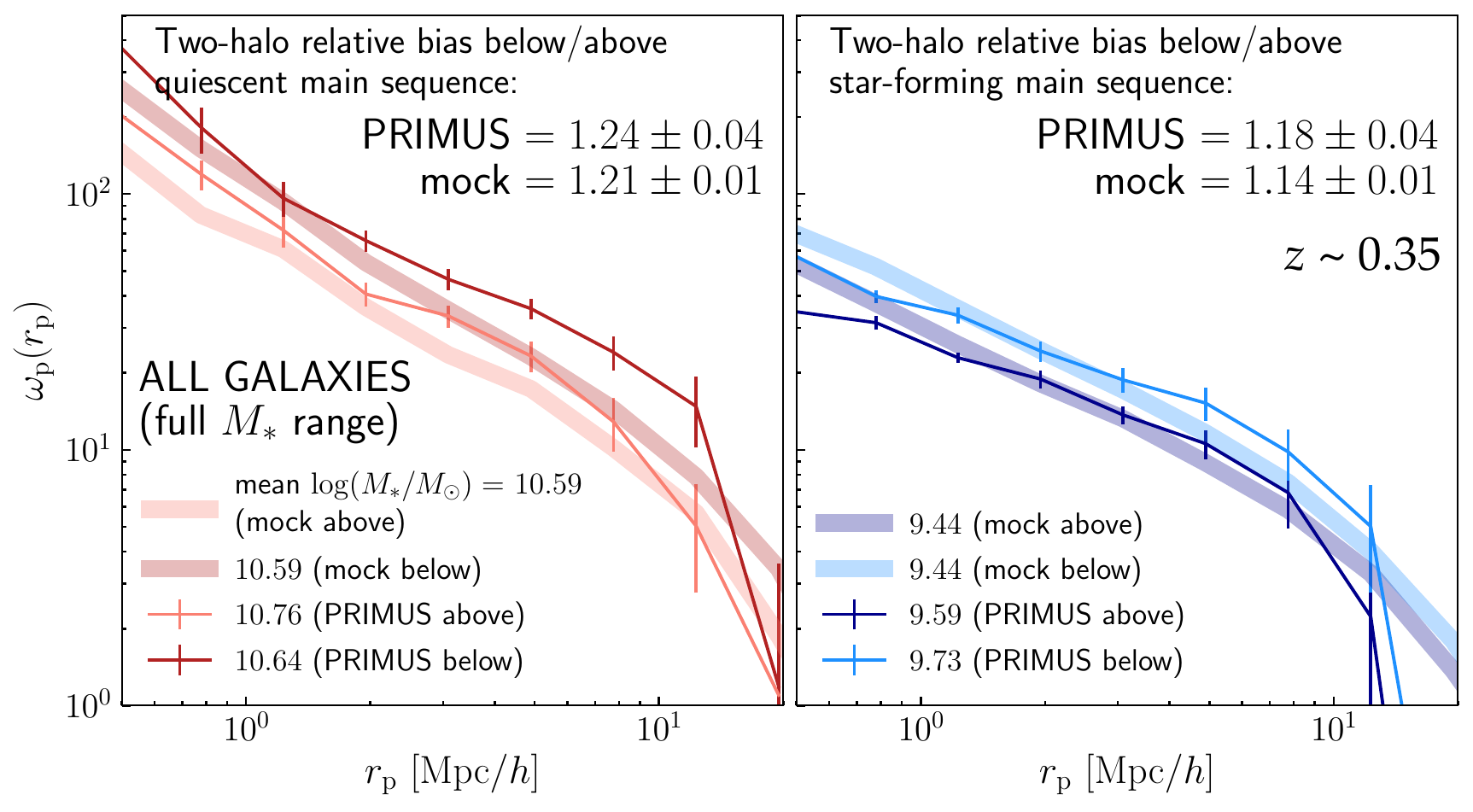}
\caption{
Clustering amplitude, \wprp, for main-sequence split samples of \emph{all} PRIMUS galaxies and stellar mass-matched samples of all galaxies from the ``modified" mock galaxy catalog. The left panel shows PRIMUS and mock galaxy samples above (thin and thick light red lines) and below (thin and thick dark red lines) the quiescent main sequence, while the right panel shows PRIMUS and ``modified" mock samples above (thin and thick dark blue lines) and below (thin and thick light blue lines) the star-forming main sequence.
}
\label{fig:wp_mssplit_all}
\end{figure*}

A main goal of this work is to quantify the relative clustering strength of star-forming and quiescent IP galaxies as a proxy for star-forming and quiescent central galaxies at $0.2<z<0.9$. As shown in the top panels of Figure~\ref{fig:wp_sfq}, quiescent IP galaxies are more strongly clustered than star-forming IP galaxies at fixed stellar mass in both redshift bins ($z\sim0.35$ and $z\sim0.7$). The top panels show \wprp for the star-forming/quiescent split PRIMUS IP galaxy samples at low ($0.2 < z < 0.5$) and high ($0.5 < z < 0.9$) redshift, and report the relative bias between quiescent and star-forming IP galaxies in each redshift bin. At low redshift the mean stellar mass of both quiescent and star-forming IP galaxies in the data is $\sim10.6$ and the relative bias is $1.71\pm0.07$, while at high redshift ($0.5<z<0.9$) the mean stellar mass is $\sim11.0$ and the relative bias is $1.31\pm0.16$.

Because IP galaxies in PRIMUS are only a proxy for true central galaxies, it is important to consider the extent to which the relative bias we measure for IP galaxies may differ from the value for  galaxies in PRIMUS that are truly centrals. Any sample of IP galaxies selected using isolation criteria will have some degree of contamination (see \S\ref{sec:results_comp_contam} below) from satellite galaxies misclassifed as isolated galaxies. One of the most robust results from HOD studies is that satellite galaxies are more strongly clustered than central galaxies of the same stellar mass; for example, the minimum mass for a dark matter to host a satellite galaxy is typically $\sim20$ times larger than the minimum halo mass required to host a central \citep[e.g.,][]{zehavi_etal05, leauthaud_etal12}. A similarly robust result deriving from empirical modeling fits to galaxy clustering is that quiescent galaxies are more strongly clustered than star-forming galaxies at fixed stellar mass \citep[e.g.,][]{tinker_etal13}.

Due to the combination of these effects working in concert, we must consider the possibility that the difference in clustering amplitude we observe between quiescent and star-forming PRIMUS IP galaxies is due not to an inherent dependence of {\it central} galaxy clustering strength on sSFR, but is instead a result of contamination by satellite galaxies when selecting IP galaxy samples.

We investigate this possibility by measuring the relative bias of quiescent versus star-forming galaxies for samples of IP and central galaxies, with the same stellar mass distribution as PRIMUS IP galaxies, from the ``standard" mock galaxy catalogs. IP galaxies are selected by applying to the mock catalogs the same isolation criteria used to select IP galaxies in PRIMUS. Table~\ref{tab:mock_samples} presents details of the mock IP and central galaxy samples, and clustering amplitudes and relative biases are shown in the bottom panels of Figure~\ref{fig:wp_sfq}.

At low redshift ($z\sim0.35$; bottom left panel) the relative bias of quiescent versus star-forming mock IP galaxies (thick red and blue lines) at fixed stellar mass is $1.38\pm0.03$, while for mock central galaxies (dashed red and blue lines) with the same stellar mass distribution the relative bias is $1.28\pm0.02$, or 7\% less than IP galaxies. Thus while the relative bias for mock central galaxies is slightly lower than for mock IP galaxies, quiescent mock central galaxies are more strongly clustered than star-forming mock central galaxies at fixed stellar mass (for the stellar mass range probed here). Applying this 7\% ``correction" to the relative bias of PRIMUS IP galaxies yields an estimated value of $1.59\pm0.06$ for central galaxies in PRIMUS at $z\sim0.35$.

We find similar results at high redshift (bottom right panel of Figure~\ref{fig:wp_sfq}), where the relative bias of quiescent versus star-forming mock galaxies is $1.36\pm0.05$ for mock IP galaxies, and $1.29\pm0.02$ for mock central galaxies. Applying this 5\% difference results in an estimated relative bias for PRIMUS central galaxies at high redshift of $1.24\pm0.12$.

The simplest explanation of these results is distinct stellar-to-halo mass relations for star-forming and quiescent central galaxies, which \citet{behroozi_etal18} also hints at for $z\gtrsim1$.
However we note the possibility that star-forming and quiescent central galaxies could have the same mean stellar mass at fixed halo mass, and that the scatter in the SMHM relation in combination with the strong positive correlation between quiescent fraction and halo mass \citep{mandelbaum_etal16}, could imply that for a sample of central galaxies at fixed {\em stellar} mass the quiescent centrals have a larger mean halo mass than star-forming centrals.

\subsection{Main sequence split}\label{sec:mssplit}

\begin{figure*}[t]
\centering
\includegraphics[width=0.8\linewidth]{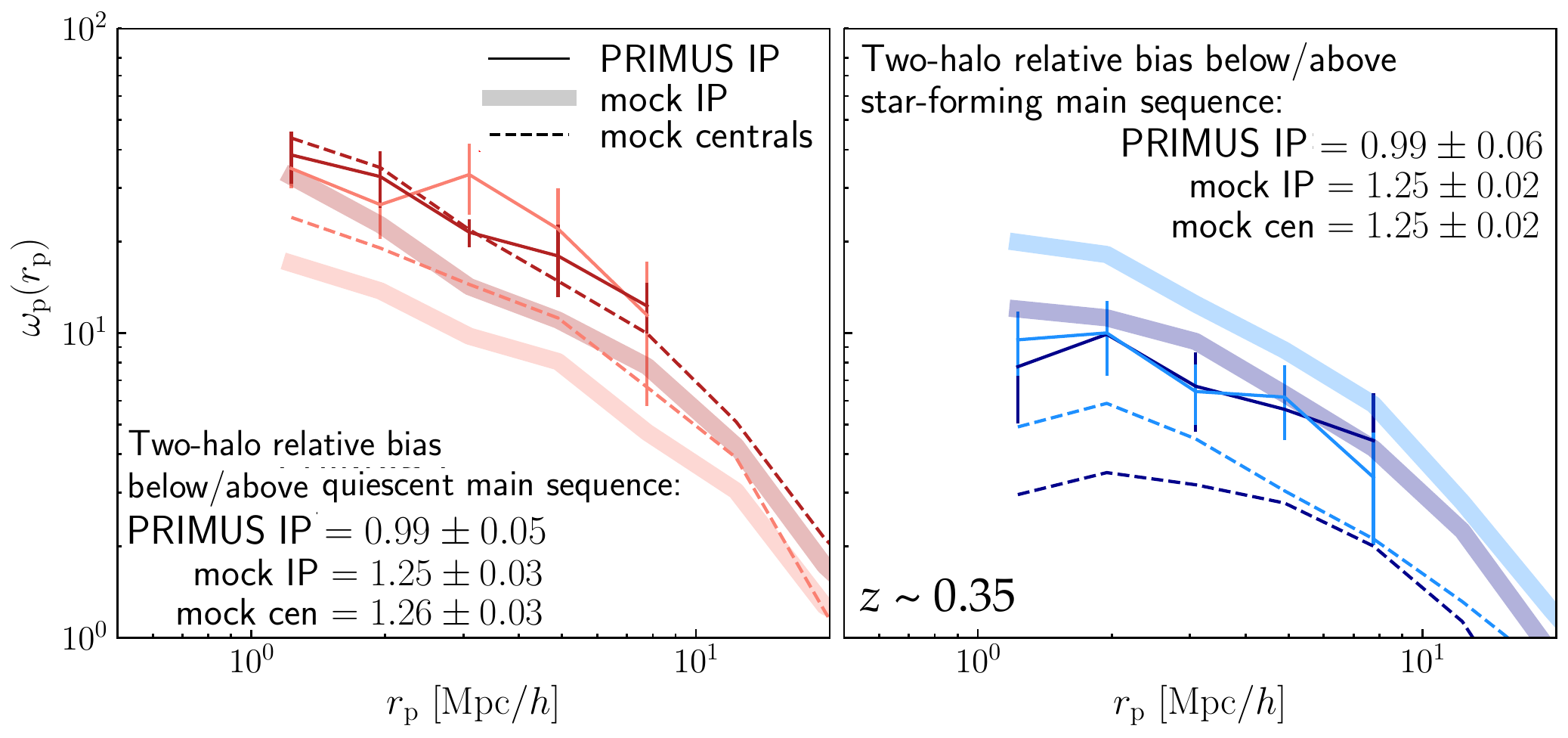}
\caption{
Clustering amplitude, \wprp, for main-sequence split samples of PRIMUS IP galaxies (solid lines), and IP galaxies (thick lines) and central galaxies (dashed lines) in the ``modified" mock galaxy catalog.
IP galaxy samples above and below the quiescent main sequence (light and dark red lines) have the same stellar mass distribution ($\logm\sim10.8$), as do the IP galaxy samples above and below the star-forming main sequence (dark and light blue lines; $\logm\sim10.3$).
The joint stellar mass and sSFR distribution of each IP galaxy sample is shown in Figure~\ref{fig:primus_ms_cuts}.
Intra-sequence relative bias is \emph{not} present for PRIMUS IP galaxies---i.e.~we don't see clustering strength dependence on sSFR for IP galaxies \emph{within} either the star-forming or quiescent main sequence. However, intra-sequence relative bias does persist for mock IP and central galaxies in the ``modified" mock galaxy catalog.
}
\label{fig:wp_mssplit}
\end{figure*}

\begin{figure*}[bt]
\centering
\includegraphics[width=0.8\linewidth]{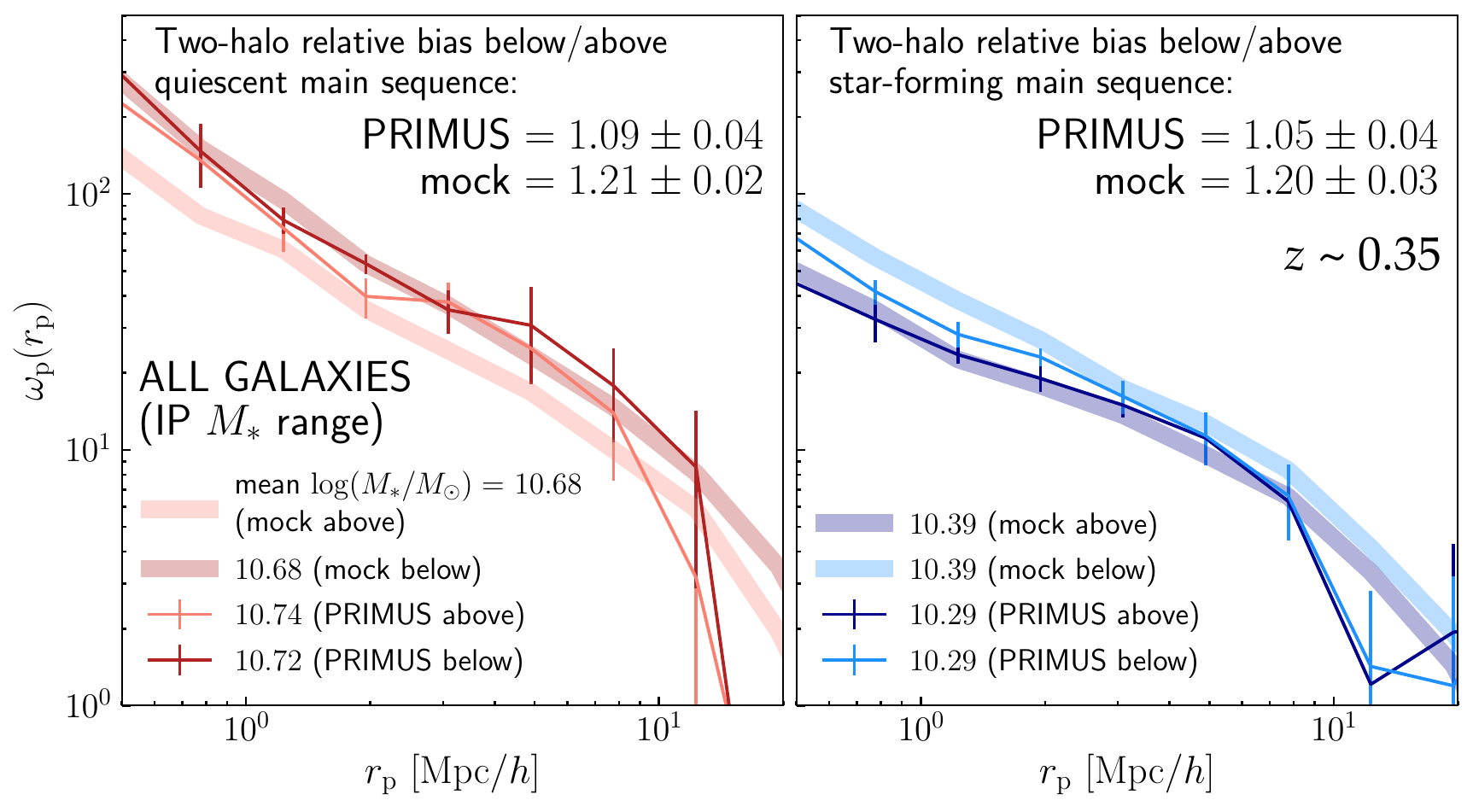}
\caption{
Same as Figure~\ref{fig:wp_mssplit_all} except here the stellar mass range of each galaxy sample is limited to that of quiescent (left) and star-forming (right) PRIMUS IP galaxies. The clustering amplitude and intra-sequence relative bias of all galaxies in the ``modified" mock galaxy catalog matches PRIMUS \emph{over the full PRIMUS stellar mass range}. This agreement is no longer present when the same galaxy samples are constrained to the (higher) PRIMUS IP galaxy stellar mass range.
}
\label{fig:wp_mssplit_all_highmass}
\end{figure*}

In the previous section we found a significant ($10\sigma$) difference in the clustering amplitude of quiescent versus star-forming IP galaxies in PRIMUS at $z\sim0.35$ (for $\logm\sim10.6$), and a $\sim2\sigma$ difference at $z\sim0.7$ (for $\logm\sim11.0$). We now investigate whether the dependence of the clustering amplitude on sSFR that we find when comparing quiescent and star-forming IP galaxies in PRIMUS persists \emph{within} the star-forming and quiescent main sequences separately. As before we compare the relative biases of IP galaxy samples in PRIMUS to the corresponding IP and central galaxy samples from a mock galaxy catalog.

Since we are now dividing PRIMUS IP galaxies into four samples---above and below each of the star-forming and quiescent main sequences---we use a single wider redshift bin ($0.2<z<0.7$) instead of the narrower low and high redshift bins used in \S\ref{sec:sfqsplit}, to increase the sample sizes. Additionally, as we are now investigating whether the clustering amplitude of IP galaxies \emph{within} each main sequence depends on sSFR at fixed stellar mas, we  no longer require the same stellar mass distribution between samples of star-forming and quiescent galaxies; we only need to constrain intra-sequence samples to have the same stellar mass distributions.  This enables us to use a larger fraction of the star-forming PRIMUS IP galaxy population than we could for the ``star-forming/quiescent split" comparison, across a wider stellar mass distribution.

As the ``standard" mock galaxy catalogs we use in the previous section do not incorporate the dependence of clustering amplitude on sSFR that \citet{coil_etal17} find at fixed stellar mass for all galaxies \emph{within} the star-forming and quiescent main sequences, in this section we use the ``modified" mock galaxy catalog for comparison with PRIMUS. As described in \S\ref{sec:mocks}, the ``modified" mock catalog has an ``intra-sequence" clustering amplitude dependence on sSFR for all galaxies intentionally built into the mock catalog, such that the ``intra-sequence" relative bias---the relative bias of galaxies below versus above the star-forming or quiescent main sequence---agrees with what \citet{coil_etal17} find for PRIMUS data. 

This agreement is shown in Figure~\ref{fig:wp_mssplit_all}, which shows \wprp for the ``main sequence split" samples of \emph{all} galaxies in both PRIMUS and the ``modified" mock galaxy catalog. The left panel of Figure~\ref{fig:wp_mssplit_all} shows PRIMUS and mock galaxy samples above (light red) and below (dark red) the quiescent main sequence, while the right panel shows PRIMUS and mock samples above (dark blue) and below (light blue) the star-forming main sequence. The intra-sequence relative bias of all quiescent galaxies is $1.24\pm0.04$ for PRIMUS and $1.21\pm0.01$ for the ``modified" mock galaxies. For all star-forming galaxies the intra-sequence relative bias is $1.18\pm0.04$ for PRIMUS and $1.14\pm0.01$ for the "modified" mock galaxies.\footnote{We note that the mean stellar masses of the galaxy samples in Figure~\ref{fig:wp_mssplit_all} differ between PRIMUS and the mock galaxy catalog by up to $\sim0.3$~dex. This is because the clustering amplitude and intra-sequence relative bias measurements for the PRIMUS samples \emph{in this figure} (all galaxies at $0.2<z<0.7$ in the full PRIMUS stellar mass range divided into samples above and below the star-forming and quiescent main sequences) are taken from \citet{coil_etal17} and contain additional galaxies from DEEP2. The ``modified" mock galaxy catalog we use here for comparison with PRIMUS is matched to only PRIMUS data, not to the slightly different combined dataset of PRIMUS and DEEP2 galaxies used by \citet{coil_etal17}.}

With agreement between PRIMUS and the ``modified" mock galaxy catalog established for all galaxies, we next investigate whether the intra-sequence relative biases persist for IP galaxies, used as a proxy for central galaxies.
The results are shown in Figure~\ref{fig:wp_mssplit}, which shows \wprp for the ``main sequence split" samples of PRIMUS IP galaxies (top panels), as well as for stellar mass-matched samples of IP and central galaxies from the ``modified" mock galaxy catalog (bottom panels). We do not find an ``intra-sequence" clustering amplitude dependence on sSFR (the ``signal") for IP galaxies:~the intra-sequence relative bias of PRIMUS IP galaxies is $0.99\pm0.05$ for quiescent and $0.99\pm0.06$ for star-forming galaxies.

However, in the ``modified" mock galaxy catalog an intra-sequence clustering amplitude dependence on sSFR {\it is} present for both IP and central galaxies, 
with a relative bias of $\sim1.25\pm0.03$ for both star-forming and quiescent mock IP and central galaxies.

Given the intra-sequence relative bias detected for mock IP galaxies, we might expect a similar signal for IP galaxies in PRIMUS. However we note that the ``modified" mock galaxy catalog is intentionally designed to have intra-sequence relative bias present at all stellar masses for the sake of comparison with PRIMUS data; that intra-sequence relative bias persists in the mock catalog at higher stellar mass does not mean we should necessarily expect to observe a signal for the same stellar mass range in PRIMUS.

As Figure~\ref{fig:wp_mssplit} shows, no such signal is observed in PRIMUS data, though this could be due to the PRIMUS mass completeness limits (see \S\ref{sec:smcl}) necessarily constraining IP galaxy samples to relatively high stellar mass ($\logm\gtrsim10.8$ for quiescent and $\gtrsim10.3$ for star-forming galaxies). This is illustrated in Figure~\ref{fig:wp_mssplit_all_highmass}, which is the same as Figure~\ref{fig:wp_mssplit_all} except the stellar mass range of each ``main sequence split" galaxy sample is now constrained to the higher stellar mass range of the corresponding IP galaxy sample.
While Figure~\ref{fig:wp_mssplit_all} shows the clustering amplitude and relative bias of ``main sequence split" samples of all galaxies over the full PRIMUS stellar mass range, Figure~\ref{fig:wp_mssplit_all_highmass} shows the same measurements for these samples \emph{limited to a higher mean stellar mass}---the same stellar mass range as the ``main sequence split" IP galaxy samples.

As discussed above, the clustering amplitude and relative bias of ``main sequence split" samples of \emph{all} galaxies agrees between the PRIMUS data and the ``modified" mock galaxy catalog when sample galaxies are selected \emph{from the full PRIMUS stellar mass range}, as seen in  Figure~\ref{fig:wp_mssplit_all}. 
These intra-sequence relative biases remain consistent in the mock catalog when sample galaxies are constrained to the higher stellar mass range of PRIMUS IP galaxies ($\logm\sim10.7$ for quiescent and $\sim10.3$ for star-forming galaxies). As shown in Figure~\ref{fig:wp_mssplit_all_highmass}, the intra-sequence relative biases for all ``modified" mock galaxies at the IP galaxy stellar mass range is $1.21\pm0.02$ for quiescent and $1.20\pm0.03$ for star-forming galaxies.  However, in the PRIMUS data these intra-sequence relative biases are only $1.09\pm0.04$ for all quiescent and $1.05\pm0.04$ for all star-forming galaxies, for the IP galaxy stellar mass range.

The fact that we do not observe a significant intra-sequence relative bias for PRIMUS IP galaxies does not necessarily indicate that no signal exists, although it does indicate that a substantial component of the signal \citet{coil_etal17} find for all PRIMUS galaxies is likely due to satellites.
This result is also consistent with \citet{zu_etal17}, who find a hint of intra-sequence clustering differences in SDSS, although measurement uncertainties in their summary statistic are too large to conclusively quantify the true strength of the correlation.

In addition to the possible stellar mass-dependent effect of this signal, it is possible that our sample sizes are simply too small for a significant detection. IP galaxies are a subset of all galaxies in a given stellar mass range.
In PRIMUS we find that the intra-sequence relative bias decreases in magnitude as the galaxy sample size also decreases from all galaxies ($N\sim22,000$ for star-forming and $N\sim6,000$ for quiescent galaxy samples) to all high mass galaxies ($N\sim10,000$ for star-forming and $N\sim5,000$ for quiescent) to IP galaxies ($N\sim6,000$ for star-forming and $N\sim3,000$ for quiescent), and for the smallest subset of PRIMUS ``main sequence split" galaxies we detect no intra-sequence relative bias at all. In other words, the PRIMUS galaxy samples for which we measure no intra-sequence relative bias are 45\% to 75\% smaller than the galaxy samples where we do detect a signal. 

We note that in the ``modified" mock galaxy catalog all galaxy sample sizes are substantially larger than the corresponding PRIMUS galaxy sample. Even the smallest mock galaxy samples (``main sequence split" IP galaxies) are at least as large as the largest PRIMUS samples, and for quiescent galaxies the smallest mock samples are nearly twice the size of the largest PRIMUS samples.

\subsection{Effect of Galaxy Number Density and PRIMUS Redshift Error}\label{sec:density_test}

The mock galaxy catalogs we use are randomly down-sampled to have the same galaxy number density as the mean density of the two largest PRIMUS fields at the redshift of each mock catalog (see \S\ref{sec:mock_match}). However, the galaxy number density of PRIMUS is a strong function of redshift, as it is a flux-limited survey. To test how galaxy number density affects IP galaxy selection and the relative bias of mock IP and central galaxies mass we create ``high" and ``low" density versions of our $z=0.35$ and $z=0.7$ "standard" mock catalogs.
High density $z=0.35$ and $z=0.7$ mock catalogs are matched to the mean galaxy number density of PRIMUS at ${0.2<z<0.25}$ and ${0.5<z<0.55}$, and are 47\% and 31\% denser than the mean density catalogs, respectively.
Low density $z=0.35$ and $z=0.7$ mocks are matched to the density of PRIMUS at ${0.45<z<0.5}$ and ${0.85<z<0.9}$, and are 44\% and 53\% less dense than the mean density mock catalogs, respectively.

We find that our results are not substantially affected by changing the galaxy number density in the mock galaxy catalogs. At low redshift ($z\sim0.35$) the relative bias of quiescent to star-forming mock IP galaxies decreases by just 1\% for the low density mock catalog and increases by 1\% for the high density mock catalog.
For mock central galaxies the relative biases changes by less than a percent for both the low and high density mock catalogs.

Additionally, the uncertainties of these low and high density relative bias measurements are not substantially different from the mean density mock catalog: $17\sigma$ and $8\sigma$ for IP galaxies in the low and high density mock catalogs, respectively, and $9\sigma$ and $17\sigma$ respectively for central galaxies in the low and high density mock catalogs.

Our results are similarly insensitive to galaxy number density at high redshift ($z\sim0.7$). The relative bias of mock IP galaxies increases by 2\% ($3.6\sigma$) and for the low density case is unchanged for the high density mock catalog ($5.5\sigma$), while for central galaxies the relative bias increases 11\% ($3.1\sigma$) and 6\% ($6\sigma$), respectively, for the low and high density mock catalogs.

We also test the effect that PRIMUS redshift errors have on the relative bias of mock IP and central galaxies by comparing to results created from the ``standard" mock galaxy catalogs \emph{without} added PRIMUS redshift errors. For this test we follow the procedure described in \S\ref{sec:mock_match} but ignore the third and final step of adding PRIMUS-like redshift-space distortions to the mock catalogs. We then select IP galaxies using the same isolation criteria we use to identify IP galaxies in the ``standard" mock catalogs (with added redshift errors), and measure the relative bias of these new IP galaxy samples that do not have redshift-space distortions.

Our results are not substantially different:~the relative bias of quiescent versus star-forming mock IP galaxies selected in mock catalogs without PRIMUS redshift errors decreases by 9\% ($9\sigma$) at low redshift and increases by 2\% ($2.3\sigma$) at high redshift. For mock central galaxies the relative bias in mock catalogs without PRIMUS redshift errors increases by 4\% ($6.3\sigma$) and 8\% ($2.9\sigma$), respectively, at low and high redshift.

These tests show that the relative biases in the mock galaxy catalogs are not sensitive to galaxy number density or to redshift errors of the magnitude present in PRIMUS data.

%% file: results_comp_contam.tex
\section{Effect of Isolation Criteria on Completeness and Contamination}\label{sec:results_comp_contam}

\begin{figure*}[tb]
\centering
\includegraphics[width=0.5\linewidth]{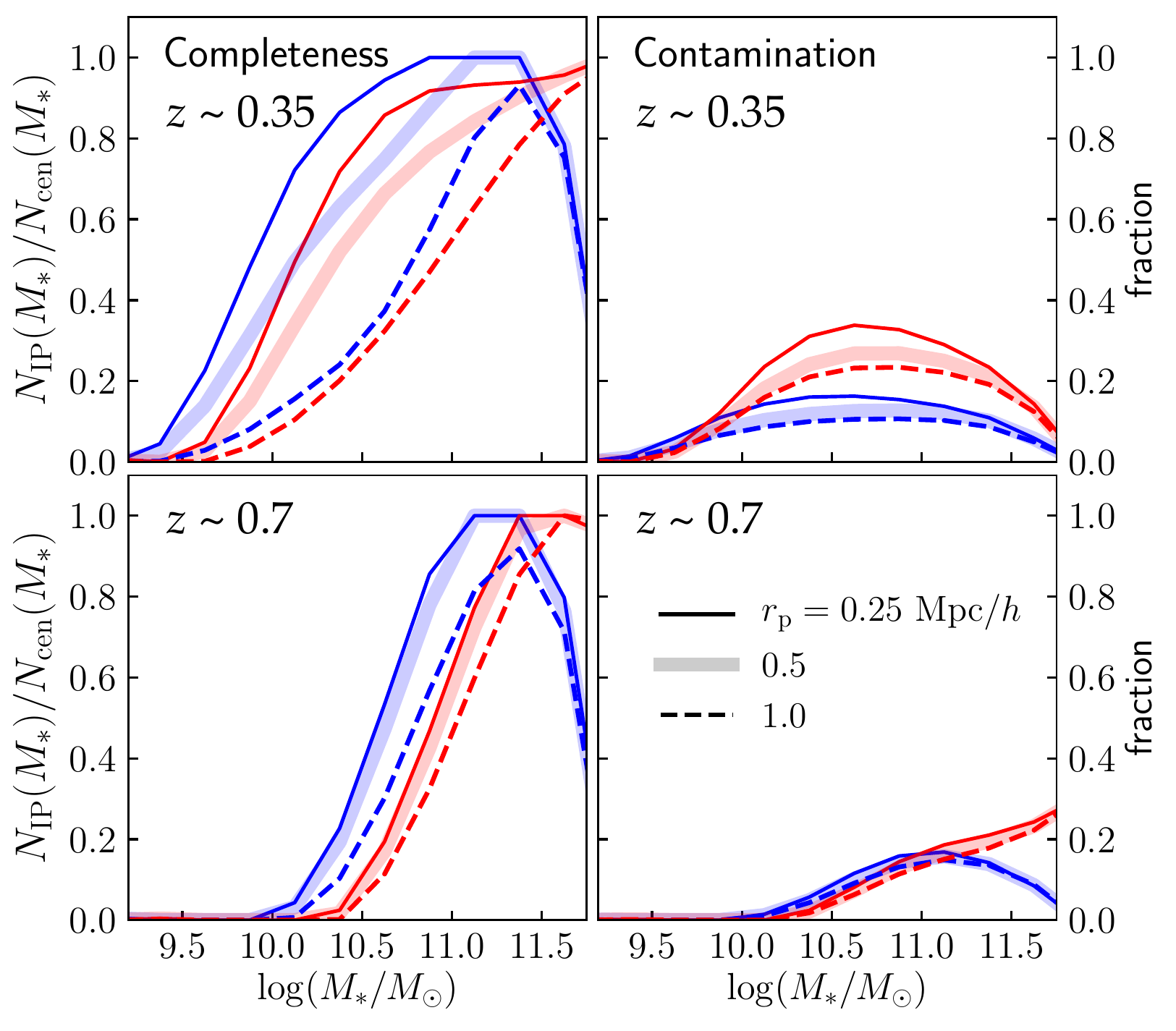}\includegraphics[width=0.5\linewidth]{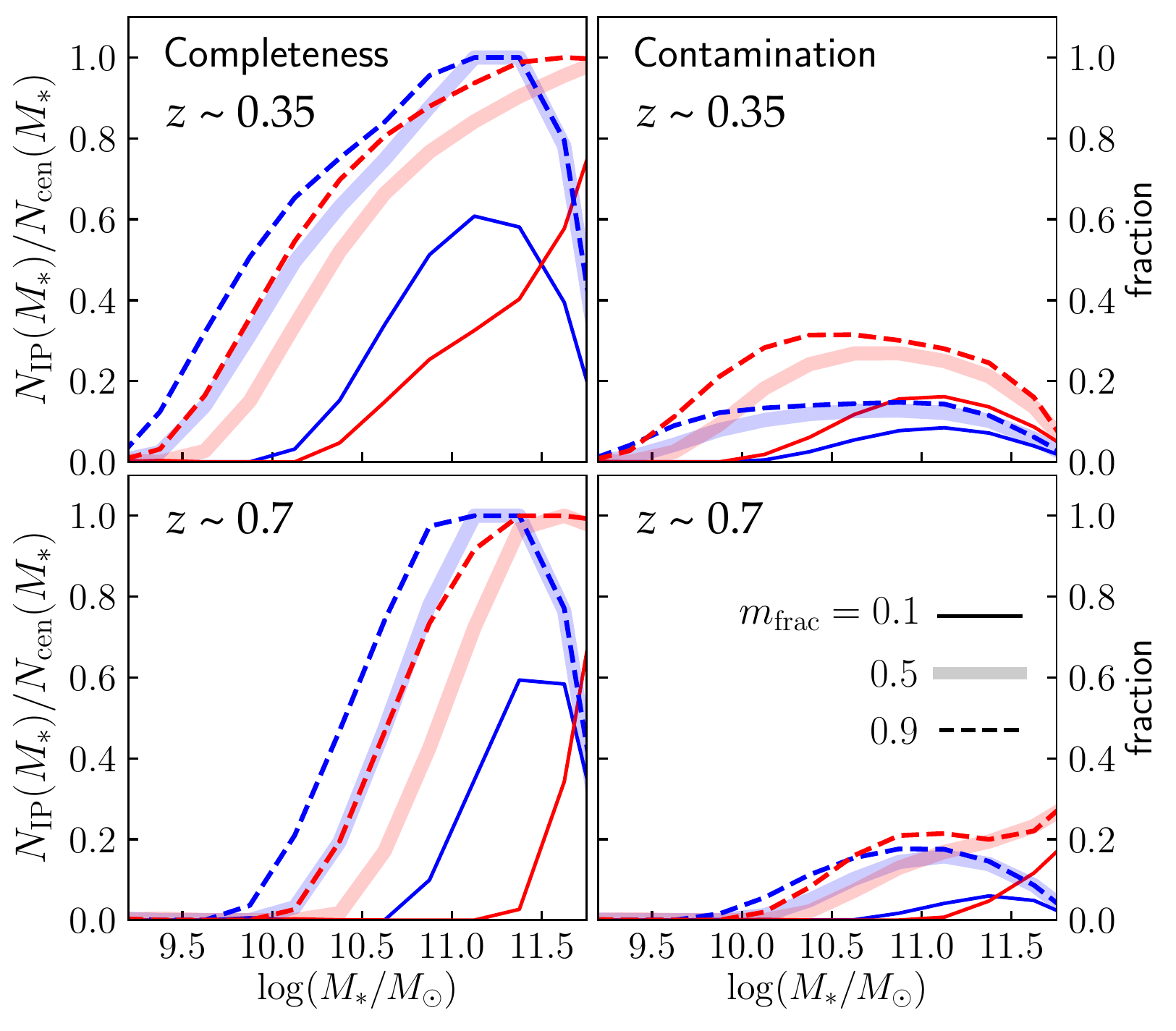}
\caption{
Completeness and contamination fractions of IP galaxy samples as a function of stellar mass for star-forming (blue) and quiescent (red) IP galaxy samples from the standard mock galaxy catalogs at $z=0.35$ and $z=0.7$ for $\rp=0.25$, 0.5, and 1.0~Mpc$/h$ (left four panels) and $\mfrac=0.1$, 0.5, and 0.9 (right four panels). IP galaxies are defined as those galaxies with no neighbors of mass ${\mstar > \mfrac\times{\rm (IP~stellar~mass)}}$ within a projected radius $r_{\rm p}$ and within a line-of-sight distance of $\pm2\sigma_{r_z}(z)$ (equivalent to $\pm2\sigmaz$, or twice the PRIMUS redshift uncertainty). Different line styles show results for different values of \rp (left) and \mfrac (right).
}
\label{fig:comp_contam}
\end{figure*}

Our goal in studying isolated primary galaxies in PRIMUS is to select a representative, unbiased sample of true central galaxies. However, as discussed in \S\ref{sec:sfqsplit}, the application of isolation criteria to the galaxy selection function introduces (unwanted) complications such as satellite contamination and biases in the distribution of environments sampled by the true centrals identified as IP galaxies.
To address these effects, in this section we use mock galaxy catalogs to quantify how both sample completeness and contamination are affected by the choice of isolation criteria used to select IP galaxies in PRIMUS.
We also demonstrate how isolation criteria yield biased galaxy samples that are not fully representative of the central galaxy population for which they are intended to be an observational proxy.

\subsection{Completeness and Contamination}\label{sec:comp_contam}

In \S\ref{sec:ip_selection} above we describe the various parameters used in the cylindrical isolation criteria we use to select IP galaxies:~projected radius (\rp), cylinder depth, and mass fraction (\mfrac), or the maximum stellar mass any galaxy within the cylinder can have for the IP candidate galaxy to be considered isolated (as a fraction of the IP galaxy candidate stellar mass). Each of these three parameters affects both the completeness and contamination of an IP galaxy sample, which in turn can affect any conclusions drawn about IP galaxies from that sample.
Completeness (measured as a function of stellar mass and galaxy type) is defined as the fraction of central galaxies identified as IP galaxies:
\begin{equation}
f_{\rm comp}(\mstar) \equiv \frac{N_{\rm (IP~\&~cen)}}{N_{\rm cen}}.
\end{equation}
\noindent Contamination is defined as the fraction of galaxies identified as IP galaxies which are actually satellite and \emph{not} central galaxies:
\begin{equation}
f_{\rm contam}(\mstar) \equiv \frac{N_{{\rm (IP~\&~{sat})}}}{N_{\rm IP}}.
\end{equation}
\noindent Ideally an IP sample will have high completeness and low contamination.

To assess how \rp, \mfrac, and cylinder depth affect the completeness and contamination of our IP galaxy samples, we reselect IP galaxies in the ``standard" mock galaxy catalogs multiple times, varying each parameter over a range of values while holding the other parameters constant. We then compute the completeness and contamination of each sample as a function of stellar mass separately for  star-forming and quiescent mock IP galaxy samples. The results are shown in Figure~\ref{fig:comp_contam} for $\rp=0.25$, 0.5, and 1.0 Mpc$/h$ (at $\mfrac=0.5$; left panels), and $\mfrac=0.1$, 0.5, and 0.9 (at $\rp=0.5$~Mpc$/h$; right panels), for a cylinder half-depth of $2.0\sigmaz$. Blue and red lines show star-forming and quiescent IP galaxies, respectively. We also test but do not show how the choice of cylinder half-depth affects completeness and contamination; we find no substantial differences over the stellar mass range of interest for cylinder half-depths between $1.0\sigmaz$ and $3.0\sigmaz$.

Ideally we would maximize completeness and minimize contamination to obtain IP galaxy samples which are maximally pure {\it and} represent the largest possible sample of central galaxies. This assumes that isolation criteria select a representative subsample of all central galaxies, which we address further in \S\ref{sec:iso_crit_bias} below. In practice, while more lenient isolation criteria increase completeness, they also increase contamination. As a main goal of this work is to assess the extent to which quiescent \emph{central} galaxies are more strongly clustered than star-forming central galaxies at fixed stellar mass, reducing IP sample contamination by satellite galaxies is more important than maximizing completeness for our purposes. At the same time, completeness cannot be entirely disregarded because adequate sample sizes are essential to keep measurement uncertainty low.

To determine which value of \rp to use (left panels of Figure~\ref{fig:comp_contam}) we consider that the contamination for quiescent IP galaxies at $M_*\sim10^{10.5}$ at $z=0.35$ increases from $\sim20\%$ for \rp=1.0~Mpc$/h$ to $\sim35\%$ for \rp$=$0.25~Mpc$/h$. While the most lenient choice of \rp$=$0.25~Mpc$/h$ would yield a larger IP galaxy sample (with higher completeness, as seen in the top left panel), the choice of $\rp=0.5$~Mpc$/h$ reduces contamination nearly 10 percentage points for quiescent IP galaxies at the peak stellar mass of the sample \emph{without} substantially reducing sample completeness. We use $\rp=0.5$~Mpc$/h$ to select IP samples from the $z=0.35$ mock galaxy catalog with relatively high completeness ($>50\%$) while retaining relatively low contamination. For the $z=0.7$ mock catalog completeness and especially contamination are less sensitive to the choice of \rp due to the lower catalog density.

As the right panels of Figure~\ref{fig:comp_contam} show, both completeness and contamination are more sensitive to the choice of \mfrac than to the choice of \rp; both fractions span a wider range as \mfrac varies from 0.1 to 0.9. While the most restrictive choice of $\mfrac=0.9$ keeps the maximum contamination value below 20\% for quiescent ($<10\%$ for star-forming) IP galaxy samples at both low and high redshift, it also substantially reduces sample completeness to a maximum value of $\sim60\%$ at $M_*\sim10^{11}\msun$ for star-forming IP galaxies at $z=0.35$, and just $\sim30\%$ for quiescent IP galaxies. Completeness is even lower for the $z=0.7$ IP galaxy samples. Our choice of $\mfrac=0.5$ substantially increases the sample size while only modestly increasing contamination to levels similar to what we obtain with the choice of $\rp=0.5$~Mpc$/h$.

It should be noted that the intrinsic scatter in the stellar mass to halo mass (SMHM) relation \citep{moster_etal13, behroozi_etal13} places a floor on the obtainable level of contamination for any IP galaxy sample; the only way to eliminate contamination entirely is to use isolation criteria so conservative that completeness is also reduced to effectively zero.

By convolving the completeness and contamination fractions shown in Figure~\ref{fig:comp_contam} as a function of stellar mass the with the stellar mass distributions of the "star-forming/quiescent split" PRIMUS IP galaxy samples, we estimate the overall completeness and contamination fractions of those samples.
At low redshift ($0.2<z<0.5$) the completeness is $63.4~(\pm0.5)\%$ for quiescent IP galaxies and $71.5~(\pm0.5)\%$ for star-forming IP galaxies, and the contamination is $24.1~(\pm0.4)\%$ for quiescent IP galaxies and $11.3~(\pm0.4)\%$ for star-forming IP galaxies.
The completeness at high redshift ($0.5<z<0.9$) is $57.1~(\pm0.6)\%$ for quiescent IP galaxies and $82.9~(\pm0.6)\%$ for star-forming IP galaxies, and the contamination is $17.4~(\pm0.6)\%$ for quiescent IP galaxies and $14.5~(\pm0.7)\%$ for star-forming IP galaxies.

Below in \S\ref{sec:iso_crit_bias} we discuss the implications of these completeness and contamination levels for both our clustering results and more generally for other studies that utilize isolation criteria.

\subsection{Biases of Isolation Criteria}\label{sec:iso_crit_bias}

\begin{figure}[b!]
\centering
\includegraphics[width=\linewidth]{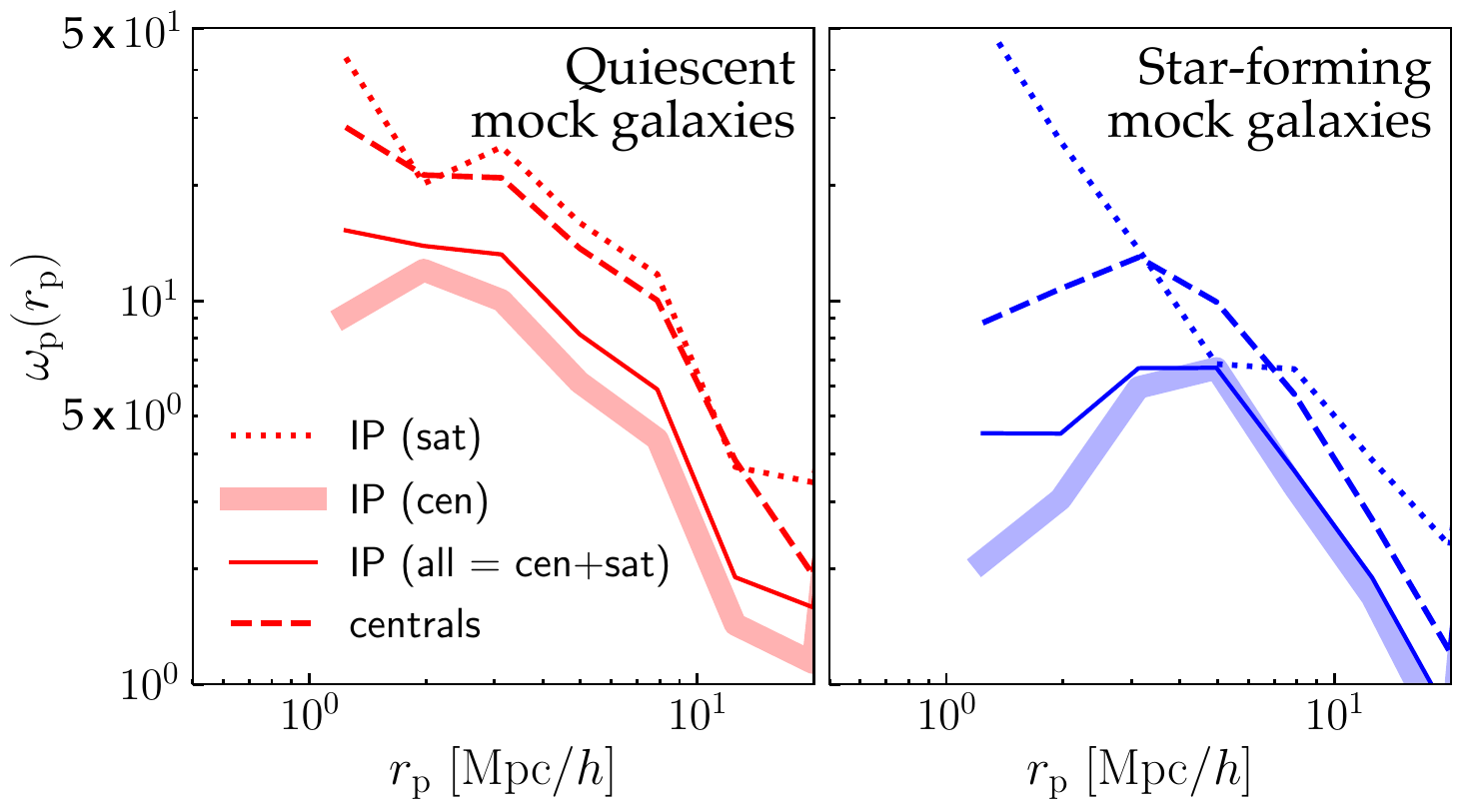}
\caption{Clustering amplitude, \wprp, for quiescent (left) and star-forming (right) samples of all mock IP galaxies (thin red and blue lines) and all mock central galaxies (thick red and blue lines) in the $z=0.35$ ``standard" mock galaxy catalog.
Also shown is \wprp for only mock IP galaxies which are actually satellite galaxies (dashed red and blue lines) and which are central galaxies (dotted red and blue lines).
At fixed stellar mass mock satellite galaxies misclassified as IP galaxies are \emph{more} clustered than all mock IP galaxies (centrals $+$ misclassified satellites), and all mock IP galaxies are \emph{more} clustered than mock IP galaxies that are central galaxies.
Mock IP galaxies that are centrals are \emph{less clustered} than all mock central galaxies at fixed stellar mass.
}
\label{fig:wp_compare_mock_IP_vs_cen}
\end{figure}

An important question when considering the contamination of an IP galaxy sample is to what extent the fraction of interloper satellite galaxies affects any scientific conclusions drawn using the sample. As discussed above, at fixed stellar mass satellite galaxies are more clustered than central galaxies, such that the significant difference we observe in the clustering strength of star-forming and quiescent IP galaxies at fixed stellar mass could be in part due to satellite contamination. We addressed this question in \S\ref{sec:sfqsplit} by using PRIMUS-like mock galaxy catalogs to quantify the difference in relative bias between mock IP and mock central galaxies at fixed stellar mass, and used this to infer the relative biases for central galaxies in PRIMUS. Even after accounting for the contributions of incompleteness and satellite galaxy contamination, we find that quiescent central galaxies are more strongly clustered at fixed stellar mass than star-forming central galaxies.  This implies that satellite contamination does \emph{not} account for the entire signal we observe for IP galaxies, and the remaining signal is a real property of central galaxies in PRIMUS data. 

\begin{figure}[b!]
\centering
\includegraphics[width=\linewidth]{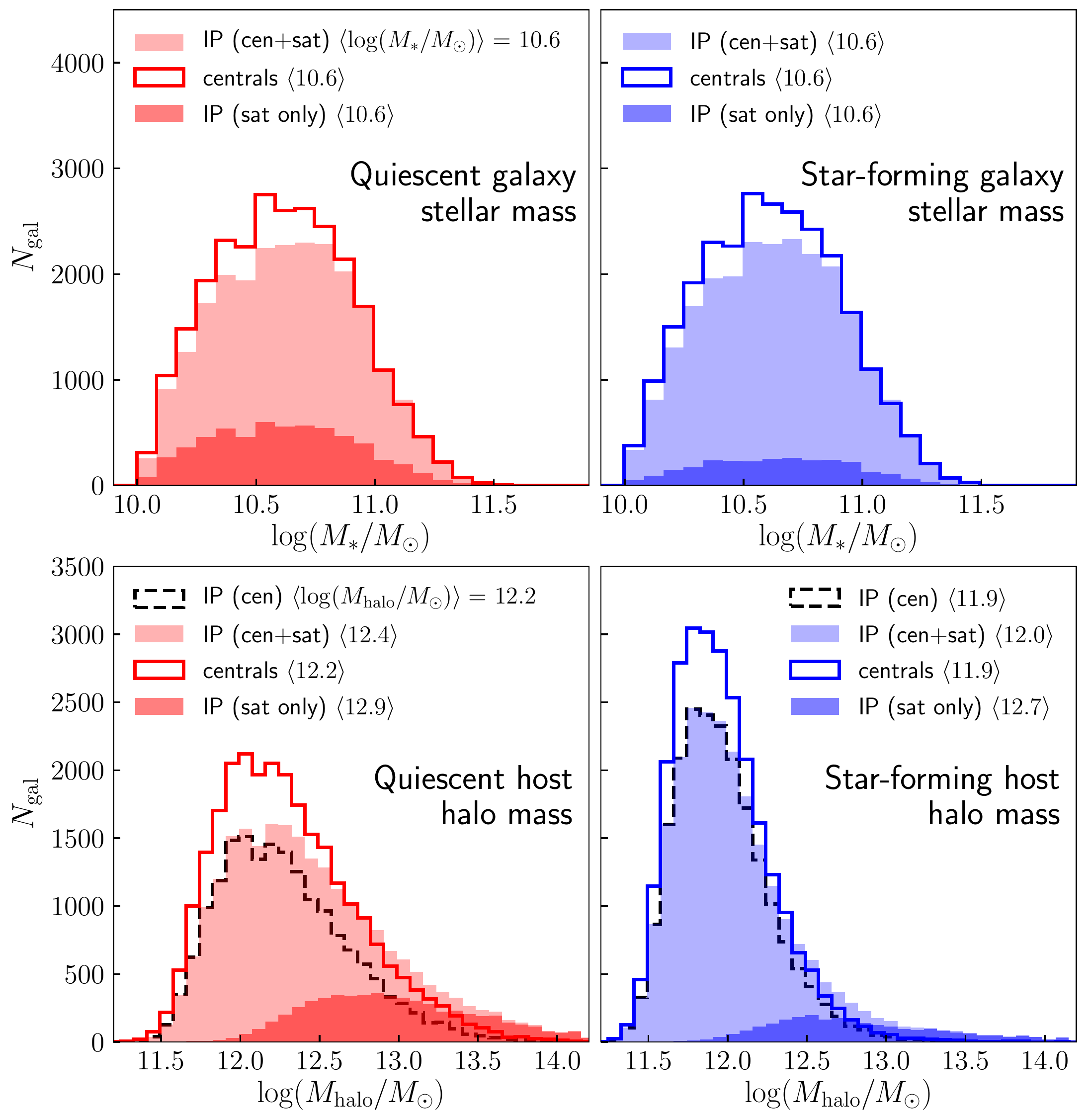}
\caption{Stellar mass (top panels) and parent halo mass (bottom panels) distributions and mean masses for the following galaxy samples in the $z=0.35$ ``standard" mock catalog:~all IP galaxies (subset of central galaxies $+$ contaminating satellites; light red and blue filled histograms), only central galaxies (red and blue unfilled histograms), and IP galaxies which are actually satellites (dark red and blue filled histograms).
The host halo mass histograms also include only IP galaxies that are also central galaxies (black dotted histograms).
}
\label{fig:mock_halo_mass_dist_lowz}
\end{figure}

After accounting for the effects of contamination, a remaining question regarding sample completeness is whether the IP galaxies identified by particular isolation criteria are a representative sample of all central galaxies, including those the isolation criteria missed, or whether some form(s) of selection bias yields an IP galaxy sample that is \emph{not} a random subsample of the central galaxies it is intended to represent.
To address this question we separate mock IP galaxies into two subsamples:~(1) IP galaxies that are also central galaxies, and (2) IP galaxies that are satellite galaxies. We then measure the two-halo clustering amplitude separately for each subsample. The results are shown in Figure~\ref{fig:wp_compare_mock_IP_vs_cen} for the following samples of quiescent and star-forming galaxies in the $z=0.35$ ``standard" mock catalog:~IP galaxies that are satellites (IP~(sat); dotted red and blue lines), IP galaxies that are true centrals (IP~(cen); thick red and blue lines), all IP galaxies (IP~(${\rm all}={\rm sat}+{\rm cen}$); thin red and blue lines), and all central galaxies (dashed red and blue lines). By design the samples of all IP galaxies and all central galaxies have the same stellar mass distributions (see top panels of Figure~\ref{fig:mock_halo_mass_dist_lowz}).

If IP galaxies were an unbiased proxy for central galaxies then we expect both samples to have the same clustering amplitude.
However, as shown in Figure~\ref{fig:wp_compare_mock_IP_vs_cen}, all IP galaxies (a subset of all centrals $+$ contaminating satellites; thick red and blue lines) are \emph{less} clustered than all central galaxies (dashed red and blue lines).

Because satellite galaxies are generically more strongly clustered than central galaxies of the same stellar mass, the presence of contaminating satellite galaxies in an IP galaxy sample should \emph{increase} the sample's clustering amplitude. This is consistent with our results:~satellite galaxies misclassified as IP galaxies (dotted red and blue lines in Figure~\ref{fig:wp_compare_mock_IP_vs_cen}) are more clustered than all IP galaxies (thin red and blue lines).

However, Figure~\ref{fig:wp_compare_mock_IP_vs_cen} also shows that all IP galaxies are \emph{less} clustered than all central galaxies at fixed stellar mass. This indicates that our IP galaxy samples are incomplete in a way that biases them towards being less clustered; i.e., isolation criteria systematically miss the most strongly clustered central galaxies.
Further, the clustering amplitude difference between all central and all IP galaxies is larger than the difference between all IP galaxies and the subset of IP galaxies that are also centrals. In other words, incompleteness has a stronger effect on the observed clustering amplitude of IP galaxies than does contamination by satellite galaxies:~the presence of misclassified satellites increases the clustering amplitude of all IP galaxies much less than the exclusion of some central galaxies decreases the it. 

The reasonable assumption of a direct mapping between clustering strength and host halo mass leads to the expectation that the host halo mass distributions of the various galaxy samples shown in Figure~\ref{fig:wp_compare_mock_IP_vs_cen} follow the same trend:~${\rm IP~(sat)} > {\rm centrals} > {\rm IP~( cen}+{\rm sat}) > {\rm IP~(cen)}$. This order of relative clustering amplitudes is summarized in Table~\ref{tab:wp_vs_halo_mass}, which also includes for comparison the same samples ordered from most to least massive mean host halo mass.
In Figure~\ref{fig:mock_halo_mass_dist_lowz} we plot histograms of the stellar mass (top panels) and host halo mass (bottom panels) distributions of the same mock galaxy samples shown in  Figure~\ref{fig:wp_compare_mock_IP_vs_cen} and Table~\ref{tab:wp_vs_halo_mass}. Because we are interested in characterizing the clustering properties of central galaxies at fixed stellar mass, all mock galaxy samples have the same stellar mass distribution and mean stellar mass (top panels of Figure~\ref{fig:mock_halo_mass_dist_lowz}) by design.

\input{wp_vs_halo_mass.tex}

We find that the host halo mass distributions of the galaxy samples in Figure~\ref{fig:wp_compare_mock_IP_vs_cen} (IP~(sat), centrals, IP~$({\rm cen}+{\rm sat})$, and IP~(cen)) do not follow the same hierarchy as the clustering amplitudes of those samples. Given that all central galaxies are more clustered than IP central galaxies at fixed stellar mass we would expect the mean halo mass of all central galaxies to be greater than that of IP central galaxies. However, the mean halo masses of all central and IP central galaxies are the same ($10^{12.2}\msun$ and $10^{11.9}\msun$ for quiescent and star-forming galaxies, respectively), indicating that the difference in clustering amplitude between all central and IP central galaxies is a function of a selection bias introduced by the isolation criteria used to select IP galaxies, and \emph{not} the result of a difference in host halo mass. We refer to the effect of this selection bias as environmental incompleteness.

\begin{figure}[b!]
\centering
\includegraphics[width=\linewidth]{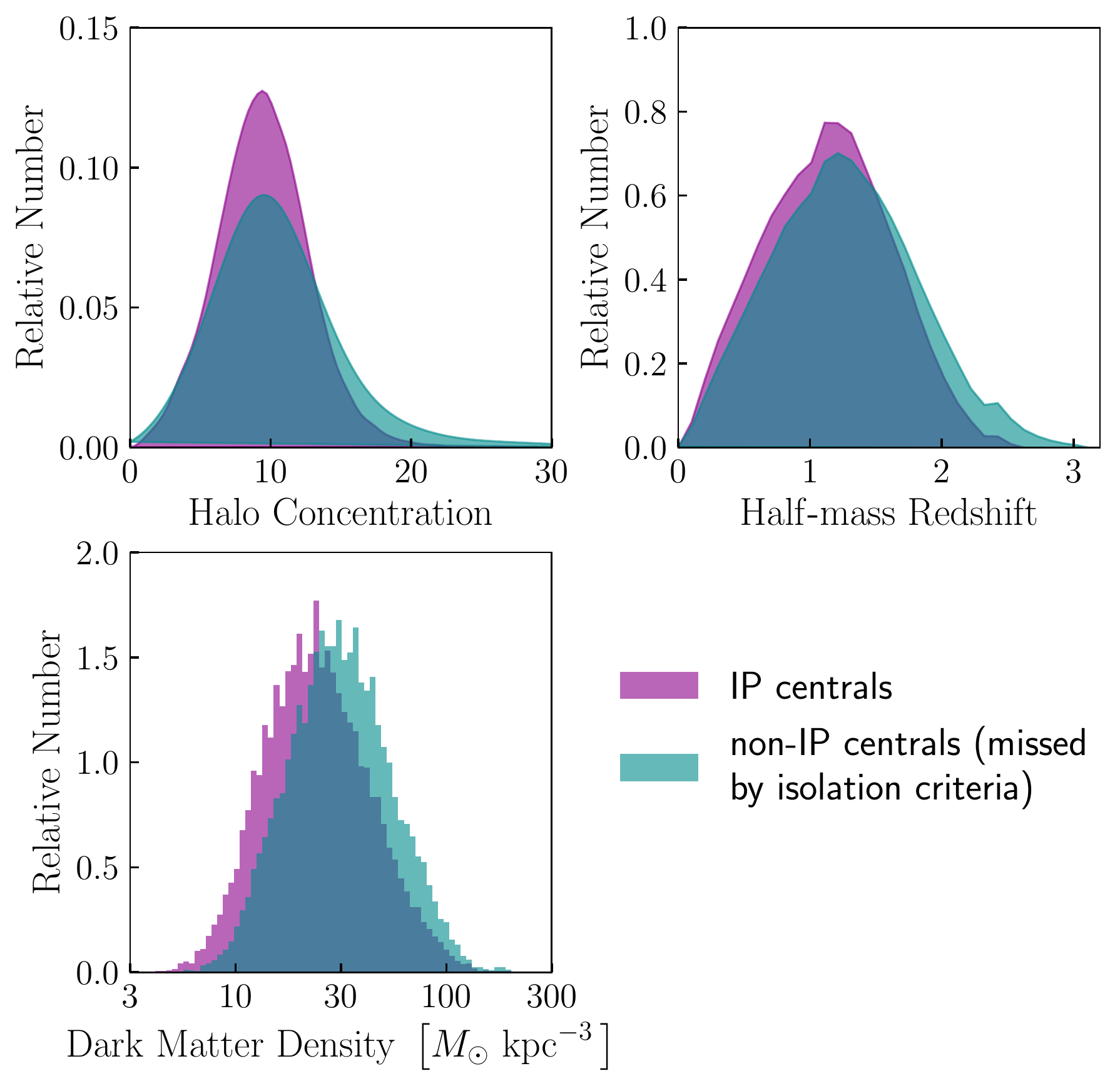}
\caption{
Distributions of mock IP galaxies that are also central galaxies (``IP centrals"; magenta) and central galaxies missed by the isolation criteria used to select IP galaxies (``non-IP centrals"; green) as a function of halo concentration (top left panel), half-mass redshift (top right panel), and large-scale dark matter density (bottom panel).
}
\label{fig:isocorrelations}
\end{figure}

As shown in Figures~\ref{fig:wp_compare_mock_IP_vs_cen} and \ref{fig:mock_halo_mass_dist_lowz}, satellite galaxies misclassifed as IP galaxies have both the largest clustering amplitude and mean parent halo mass ($10^{12.9}\msun$ and $10^{12.7}\msun$ for quiescent and star-forming satellites, respectively). In the case of all IP galaxies compared to IP satellite galaxies, IP satellite galaxies reside in substantially more massive parent halos and are more clustered than all IP galaxies. Thus contamination of an IP galaxy sample by misclassified satellite galaxies increases both the sample's mean halo mass and clustering amplitude compared to pure IP central galaxies.

In summary, satellite galaxy contamination inflates the observed clustering of IP galaxies, while (environmental) incompleteness has the opposite effect, systematically excluding central galaxies in overdense environments.
These inherent biases of isolation criteria have implications beyond the central galaxy clustering that is the focus of this work. Any study that relies on isolation criteria to identify central galaxies, such as measurements of galactic conformity \citep[e.g.,][]{kauffmann_etal13, hartley_etal15, kawinwanichakij_etal16, berti_etal17}, should not assume that such criteria generate a representative subsample of central galaxies, even after accounting for contamination by misclassified satellite galaxies.

Figure~\ref{fig:isocorrelations} provides further insight into the relationship between isolation criteria and environmental incompleteness. The three panels compare the distributions of mock IP galaxies that are also central galaxies (``IP centrals") and central galaxies missed by the isolation criteria used to select IP galaxies\footnote{Here we do not use the PRIMUS-like mock galaxy catalogs. Galaxy samples are instead selected from a $z=0$ mock catalog without added PRIMUS-like redshift errors and limited to the stellar mass range {$10.0 < \logm < 10.25$}. IP galaxies are those with no more massive galaxies within a spherical region of radius 1.5~Mpc.} (``non-IP centrals") as a function of three different parameters:~halo concentration (top left panel), half-mass redshift (top right panel), and large-scale dark matter density $\rho_{\rm DM}$ (bottom panel). Note that Figure~\ref{fig:isocorrelations} explicitly does not consider satellite galaxies and focuses entirely on distinctions between central galaxies selected versus those excluded by isolation criteria.

As shown in the upper left panel of Figure~\ref{fig:isocorrelations}, the distribution of halo concentrations of non-IP central galaxies (blue) has a long tail extending toward higher concentration that is absent from the distribution for IP central galaxies (magenta). While both distributions peak at a halo concentration of $\sim10$, isolation criteria exclude central galaxies that reside in the most highly concentrated halos. 

The upper right panel of Figure~\ref{fig:isocorrelations} shows the distributions of half-mass redshift (the redshift at which a halo had assembled half of its current virial mass) for IP central and non-IP central galaxies. The distribution for non-IP central galaxies has a similar tail toward higher half-mass redshift, corresponding to earlier halo half-mass assembly time (earlier halo formation epoch and older halo age). The IP central galaxy distribution is conversely shifted slightly toward younger halos with lower half-mass redshifts.

The bottom panel of Figure~\ref{fig:isocorrelations} shows the distributions for IP central and non-IP central galaxies of the large-scale dark matter density, defined as the spherical shell with inner and outer radii of 5 Mpc and 10 Mpc, respectively. Both distributions approximate a normal distribution, but the peaks are slightly offset to either side of {$\rho_{\rm DM}\approx30~\msun~{\rm kpc}^{-3}$}; IP centrals toward lower density and non-IP centrals toward higher density. Isolation criteria are more likely to exclude central galaxies that reside in regions of higher large-scale dark matter density beyond the scale of a single halo. 

%% file: wp_vs_halo_mass.tex
\begin{deluxetable}{cc|cc}[bt]
\tablecaption{
Summary chart of Figures~\ref{fig:wp_compare_mock_IP_vs_cen} and \ref{fig:mock_halo_mass_dist_lowz}. All central and IP central galaxies reside in host halos of the same mass, but all central galaxies are more strongly clustered than IP central galaxies due to environmental incompleteness associated with isolation criteria.
\label{tab:wp_vs_halo_mass}
}
\tablewidth{0pt}
\tablehead{
\colhead{Clustering} &
\colhead{Galaxy} &
\colhead{Host halo} &
\colhead{Galaxy} \\
\colhead{amplitude}&\colhead{sample}&\colhead{mass}&\colhead{sample}
}
\startdata
most & IP (sat) & most & IP (sat) \\
clustered & \multirow{2}{*}{centrals} &
massive & \multirow{3}{*}{IP (cen + sat)} \\
\multirow{2}{*}{$\downarrow$} & & \multirow{2}{*}{$\downarrow$} & \\
& \multirow{2}{*}{IP (cen + sat)} & & \\
least & & least & centrals \& \\
clustered & IP (cen) & massive & IP (cen) \\
\enddata

\end{deluxetable}

%% file: conclusion.tex
\section{Summary and Discussion}\label{sec:summary}

We use the PRIMUS galaxy redshift survey to investigate the extent to which the known stronger observed clustering amplitude of quiescent galaxies compared to  star-forming galaxies at a given stellar mass is due to clustering differences between central galaxies in these two populations. Using a total sample of $\sim60,000$ spectroscopic redshifts at $0.2<z<0.9$, we identify ``isolated primary" (IP) galaxies, using isolation cuts in spatial proximity and stellar mass of nearby galaxies.  We then compare the clustering amplitudes of quiescent and star-forming IP galaxies at a given stellar mass, where IP galaxies are used as an observational proxy for central galaxies.

We further test for any dependence of the clustering amplitude on the specific SFR \emph{within} each of the star-forming and quiescent main sequences (which we term ``intra-sequence" relative bias). We also compare the observed clustering of IP galaxies in PRIMUS data with the same measurements for PRIMUS-like mock galaxy catalogs created from cosmological simulations, to understand how the clustering of {\it true} centrals may differ from that of IP galaxies. We use these mock catalogs to quantify the completeness and contamination of our IP galaxy samples and demonstrate that using isolation criteria to identify central galaxies creates biased subsamples of the true central galaxy population. 

Our main results are:

\begin{enumerate}

\item{
At fixed stellar mass quiescent IP galaxies are more strongly clustered than star-forming IP galaxies in the PRIMUS dataset. The relative bias observed is 
$1.71\pm0.07$ ($10\sigma$) at $z\sim0.35$ and
$1.31\pm0.16$ ($1.6\sigma$) at $z\sim0.7$.
We use our PRIMUS-like mock catalogs to interpret the magnitude of this relative bias as being caused, in part, by quiescent {\em central} galaxies being more strongly clustered relative to star-forming centrals within these two populations. These differences are consistent with the recent halo occupation models of \citet{behroozi_etal18}, and can thus be accounted for by distinct stellar-to-halo mass relations for quenched versus  star-forming centrals, and/or by central galaxy assembly bias.
}

\item{
We do not find evidence that the clustering strength of IP galaxies depends on specific SFR \emph{within} either the star-forming or quiescent main sequence (so-called ``intra-sequence relative bias"), which \citet{coil_etal17} find for {\it all} galaxies in PRIMUS at fixed stellar mass.
However, we cannot rule out the existence of intra-sequence relative bias for IP (or central) galaxies, in part because the sample sizes decrease 
when further subdividing the star-forming and quiescent populations, and because of the limited stellar mass range over which we can perform this test.
}

\item{
We used mock galaxy catalogs to study both incompleteness (true central galaxies excluded by isolation criteria used by observers) and contamination (satellite galaxies misclassified as isolated primary galaxies). We find that both incompleteness and contamination depend significantly on stellar mass, redshift, and galaxy type (star-forming or quiescent).
In particular, the same isolation criteria applied to star-forming and quiescent galaxies yields IP galaxy samples with higher completeness for star-forming galaxies at all stellar masses and redshifts probed here.
Completeness is also a strong positive function of stellar mass for both galaxy types.
At low redshift, contamination is nearly twice as high for quiescent IP galaxies compared to star-forming IP galaxies, while the IP galaxy samples have similar levels of contamination at higher redshift.

}

\item{
Both star-forming and quiescent mock IP galaxies are less clustered than true central galaxies at fixed stellar mass, even though they reside in equally massive halos. The isolation criteria often used by observers to select ``central" (i.e. isolated) galaxies can yield biased samples by preferentially excluding true central galaxies in the most overdense regions, a phenomenon we refer to as {\em environmental incompleteness}.
While contamination by satellite galaxies enhances the observed clustering of IP galaxies, environmental incompleteness decreases the clustering amplitude.
This and other systematic biases that result from using isolation criteria to select proxies for central galaxies can be addressed by comparing observational data with mock galaxy catalogs, as is done here.
}

\end{enumerate}

One of our principal findings is that the application of isolation criteria generically influences central versus satellite contributions to galaxy clustering, as well as modulates the role of central galaxy environment on the signal. Measurements of the clustering of IP galaxies thereby provide important, distinct information that is not contained in standard observations of the two-point clustering of all galaxies. Thus our new measurements of the two-point function of quiescent versus star-forming IP galaxies should be substantially constraining for galaxy evolution models \citep[e.g.,][]{benson12, becker15, lacey_etal16, moster_etal17, henriques_etal17, behroozi_etal18}. Looking forward, models of galaxy formation would benefit from measurements of IP clustering made with larger galaxy samples that probe lower stellar masses at intermediate redshift, which are currently unavailable with existing spectroscopic redshift surveys.

The potentially insidious effects of satellite galaxy contamination have been explored previously in the literature. In particular, it is common to design and apply isolation criteria with the primary goal of minimizing satellite contamination in pursuit of a pure sample of centrals \citep[e.g.,][]{tal_etal13, planck13LBG, hartley_etal15, anderson_etal15, mandelbaum_etal16}. A basic prediction of halo occupation statistics is that satellites generically live in higher mass parent halos relative to centrals of the same stellar mass. Thus we should {\em generically} expect that satellite contamination boosts the clustering of putative centrals, which is consistent with our measurements of the relative bias of IP galaxies to true centrals. We thus confirm that contamination by satellite galaxies can indeed significantly bias attempts to measure the large-scale structure of true centrals.

On the other hand, environmental incompleteness can also substantially bias attempts to measure central galaxy clustering. Due to coupling between small and large scales, we should generically expect the application of isolation criteria to exclude centrals in overdense regions of the underlying matter density field. This is borne out in 
mock catalogs, as shown in Figure \ref{fig:isocorrelations}.
Moreover, we find that IP galaxy samples preferentially exclude true centrals in high-concentration, early-forming halos. This has particularly important implications for efforts to observationally detect galaxy assembly bias, or more generally, to measure and constrain the co-evolution between galaxies and their halos. Recent progress in understanding the physical origin of assembly bias suggests that the bulk of the underlying halo assembly bias signal is driven by only $\sim10\%$ of true host halos of a given mass \citep{mansfield_kravtsov19}. Thus when attempting to observationally constrain central galaxy assembly bias by applying isolation criteria to a galaxy sample, if sufficiently aggressive criteria are applied, one could in principle entirely erase the effect by systematically discarding the sample of dark matter halos responsible for the signal.
As shown in \citet[][see their Figures 5 and 6]{lee_etal17}, the possibility of such a bias is not limited to analyses based on the particular choice of halo property adopted in \citet{mansfield_kravtsov19}. In fact, for {\em many} choices of secondary halo property, correlations with large-scale density only exist for those halos residing in the high-density tail of the distribution of environments.

The potential pitfalls of environmental incompleteness apply not just to isolation criteria, but also to methods based on group-finding algorithms. Galaxy group catalogs offer a closely related but alternative approach to identifying samples of central galaxies, and there are many examples in the literature in which group-finding algorithms are used to exclude satellites from a subsequent measurement \citep[e.g.,][]{yang_etal05, weinmann_etal06, berlind_etal06, yang_etal08, lin_etal16}. As shown previously \citep{duarte_mamon14, campbell_etal15}, systematic errors in group-finding algorithms can lead to significant biases in the trends inferred directly from a sample of group-finder-determined centrals. Here we point out that environmental incompleteness effects are not limited to samples selected with isolation criteria; this same systematic applies to central galaxy identification with a group-finding algorithm. Similarly, measurements of galactic conformity \citep[e.g.,][]{kauffmann_etal13, hartley_etal15, kawinwanichakij_etal16, berti_etal17, tinker_etal18} should not generally assume that such criteria create a fully representative subsample of central galaxies. 

Observational studies that select subsamples dependent on spatial location from the full galaxy population should apply their selection methods to mock galaxy catalogs, created from $N$-body simulations combined with galaxy halo occupation models, to assess the completeness and contamination of their samples. Mock catalogs can additionally be useful in defining a  statistic that is robust to the dominant systematics in the problem, as we have done here by tuning our isolation criteria \citep[for another recent example see][]{calderon_etal18}.
Analyses with mock catalogs are also especially useful in determining whether galaxy samples are subject to unanticipated selection biases, as well as to address the effects of any biases by both refining selection criteria and quantifying differences between statistics obtained from real versus simulated data \citep[e.g.,][]{barton_etal07}.